\documentclass[10pt]{article}
\usepackage[pdftex]{graphicx}
\usepackage{caption}
\usepackage{subcaption}
\usepackage{dcolumn}
\usepackage{bm}
\usepackage{color}
\usepackage{amsfonts}
\usepackage{amsmath}
\usepackage{stmaryrd}
\usepackage{epsfig}
\usepackage[english]{babel}
\usepackage[all]{xy}
\usepackage{ulem}
\usepackage[top=25.4mm, bottom=25.4mm, left=31.7mm, right=32.2mm]{geometry}
\usepackage{amssymb}

\DeclareMathOperator{\arccosh}{arccosh}
\DeclareMathOperator{\arctanh}{arctanh}
\DeclareMathOperator{\sech}{sech}
\DeclareMathOperator{\arccoth}{arccoth}

\usepackage{xcolor}
\usepackage{multicol}
\usepackage{color}

\definecolor{LightBlue}{rgb}{0.8,0.8,1}

\renewcommand{\theequation}{\arabic{equation}}

\input{tcilatex}
\begin{document}

\title{\textbf{On a class of 3D second-order integrable Lagrangians  and their dispersive deformations}}
\author{ Lingling Xue$^{1}$, E.V. Ferapontov$^{2}$, M.V. Pavlov$^{1}$}
\date{}
\maketitle

\renewcommand{\baselinestretch}{1.25}

\vspace{-5mm}

\begin{center}
$^{1}$Department of Applied Mathematics\\[0pt]
Ningbo University\\[0pt]
Ningbo 315211, P.R. China \\[0pt]
\ \\[0pt]
$^{2}$Department of Mathematical Sciences \\[0pt]
Loughborough University \\[0pt]
Loughborough, Leicestershire LE11 3TU, UK \\[0pt]
\ \\[0pt]
e-mails: \\[1ex]
\texttt{xuelingling@nbu.edu.cn}\\[0pt]
\texttt{E.V.Ferapontov@lboro.ac.uk}\\[0pt]
\texttt{maksim@nbu.edu.cn}\\[0pt]
\end{center}

\bigskip

\begin{abstract}
We investigate integrability of Euler-Lagrange equations associated with 3D
 second-order Lagrangians of the form
\begin{equation*}
\int f(u_{xy},u_{xt},u_{yt})\ \text{d}x\text{d}y\text{d}t.
\end{equation*}
It is demonstrated  that there are exactly four different types of such Lagrangian densities $f$: the first one is given by the formula $f=\sqrt{u_{xy}u_{xt}u_{yt}}$, the second and the third are more complicated (although still representable in elementary functions), whereas the most generic fourth one
 is expressible in terms of the Lobachevsky function,
 revealing unexpected links  to spherical/hyperbolic trigonometry and Schl\"afly-type formulas.
 Dispersionless Lax pairs and integrable dispersive deformations of the corresponding Euler-Lagrange equations are also constructed.
 Remarkably, dispersive deformation of the first Lagrangian density coincides with the Lagrangian of the classical Darboux system arising in the theory of triply-orthogonal coordinate systems in $\mathbb{R}^3$.
 Dispersive deformations of the three other cases provide Lagrangian formulation of  semi-discrete and fully discrete versions of the Darboux system, with one, two and three discrete variables, respectively.
\end{abstract}


\bigskip


\newpage

\bigskip

\tableofcontents


\section{Introduction and summary of the main results}
\label{sec:intro}

The existing classification results of dispersionless integrable Lagrangians have demonstrated remarkable and unexpected links to the theory of special functions (modular forms, Jacobi forms, multi-dimensional hypergeometric functions), the theory of Frobenius manifolds (WDVV equations of topological field theory), and integrable conformal geometry.  Thus, 3D first-order integrable Lagrangians of the form
$$
\int f(u_{x},u_{y},u_{t})\ \text{d}x\text{d}y\text{d}t
$$
were extensively studied in \cite{FKT, FO}; it was shown in \cite{CFOZ} that the  generic integrable Lagrangian density $f$ is a Picard-type modular form of its arguments (we say that a Lagrangian density is `integrable' if  the corresponding Euler-Lagrange equation is integrable). In paper \cite{FPX}, we have classified 2D second-order integrable Lagrangians of the form
$$
\int f(u_{xx},u_{xy},u_{yy})\ \text{d}x\text{d}y,
$$
demonstrating that the integrability conditions for the Lagrangian density $f$ are equivalent, via a Legendre transformation, to  WDVV equations (note that 2D {\it first-order} Lagrangian densities $f(u_x,u_y)$ generate linearisable Euler-Lagrange equations and, therefore, are automatically `integrable').  In \cite{FPX}, we have  initiated the classification of 3D second-order integrable Lagrangians,
$$
\int f(u_{xx}, u_{xy}, u_{yy}, u_{xt}, u_{yt}, u_{tt})\ \text{d}x\text{d}y\text{d}t,
$$
however, our results in this direction are incomplete due to  complexity of the problem.

In this paper, we classify 3D second-order  Lagrangians of the special type
\begin{equation}
\int f(u_{xy},u_{xt},u_{yt})\ \text{d}x\text{d}y\text{d}t,  \label{lag}
\end{equation}
such that the corresponding Euler-Lagrange equations,
\begin{equation*}
\left( \frac{\partial f}{\partial u_{xy}}\right) _{xy}+\left( \frac{\partial
f}{\partial u_{xt}}\right) _{xt}+\left( \frac{\partial f}{\partial u_{yt}}
\right) _{yt}=0,
\end{equation*}
which are fourth-order quasilinear PDEs for $u(x,y,t)$, are integrable (in the sense to be explained in Sections \ref{sec:quasi} and \ref{sec:int}). Lagrangians of type  (\ref{lag}) are already highly non-trivial; it turns out that their integrable dispersive deformations  provide variational formulation of the classical Darboux system, including its differential-difference and fully discrete versions.

In Section \ref{sec:Lob} we derive integrability conditions for the Lagrangian density $f$ by applying the method of hydrodynamic reductions to the corresponding Euler-Lagrange equation. The integrability conditions constitute an involutive system of Monge-Amp\`ere equations (\ref{int}) for the density $f$. Solving the integrability conditions leads, modulo elementary normalisations, to the four essentially different canonical forms of integrable Lagrangian densities:

\medskip

\noindent{\bf Case 1. }
$$
f= \sqrt{u_{xy}u_{xt}u_{yt}}.
$$

\noindent{\bf Case 2. }
\begin{equation*}
\begin{array}{c}
f=u_{xt}u_{yt}\sqrt{1-\frac{2u_{xy}}{u_{xt}u_{yt}}}-2u_{xy}\arctanh \sqrt{1-\frac{2u_{xy}}{u_{xt}u_{yt}}}.
\end{array}
\end{equation*}

\noindent{\bf Case 3. }
\begin{equation*}
\begin{array}{c}
f=(u_{xt}-u_{xy})\arctanh \frac{\sqrt{u_{xt}^{2}+u_{xy}^{2}-2u_{xt}u_{xy}\coth u_{yt}%
}}{u_{xt}-u_{xy}}-(u_{xt}+u_{xy})\arctanh \frac{\sqrt{u_{xt}^{2}+u_{xy}^{2}-2u_{xt}u_{xy}\coth
u_{yt}}}{u_{xt}+u_{xy}}.
\end{array}
\end{equation*}

\noindent{\bf Case 4.}
\begin{equation*}
\begin{array}{c}
f=Au_{yt}+Bu_{xt}+Cu_{xy}+{\cal L} \left(\frac{2\pi-A-B-C}{2}\right) +{\cal L} \left(\frac{A+B-C}{2}\right) +{\cal L} \left(\frac{A+C-B}{2}\right) +{\cal L} \left(\frac{B+C-A}{2}\right)
\end{array}
\end{equation*}
where ${\cal L}(s)=-\int_0^s\ln \cos \xi \, d\xi$ is the Lobachevsky function, which was used by Lobachevsky in the computation of hyperbolic volumes. To specify the dependence of $A, B, C$ on the derivatives $u_{xy},u_{xt},u_{yt}$, consider a spherical triangle with interior angles $A, B, C$ and the opposite edge lengths $a, b, c$,  and set
$$
\cos a=\tanh u_{yt}, \quad \cos b=\tanh u_{xt}, \quad \cos c=\tanh u_{xy}.
$$
Then $A, B, C$ can be recovered from the spherical laws of cosines,
\begin{equation*}
\begin{array}{c}
\cos a=\cos b\cos c+\sin b\sin c\cos A,\\
\cos b=\cos a\cos c+\sin a\sin c\cos B,\\
\cos c=\cos a\cos b+\sin a\sin b\cos C.
\end{array}
\end{equation*}
Note that the so defined density $f$ has the meaning of `capacity'  of a spherical triangle which has appeared in the context of variational principles for circle packings and triangulated surfaces, see Section \ref{sec:Lob} for further details and references.

\medskip

In Section \ref{sec:dispLax} we calculate dispersionless Lax pairs for the Euler-Lagrange equations corresponding to  Lagrangian densities 1-4. The structure of dispersionless Lax pairs clearly indicates that these equations are commuting flows of the dispersionless KP hierarchy.

\medskip

In Section \ref{sec:dispdef} we obtain integrable dispersive deformations of the Lagrangian densities 1-4 (by constructing dispersive deformations of the corresponding dipersionless Lax pairs). Thus, in Case 1, which is already  non-trivial and interesting, the Lagrangian density $f=\sqrt{u_{xy}u_{xt}u_{yt}}$ possesses integrable dispersive deformation,
$$
F=L+\frac{\epsilon}{4}\, u_{xyt} \ln \frac{2L-\epsilon u_{xyt}}{2L+ \epsilon u_{xyt}},
$$
where $L=\sqrt{u_{xy}u_{xt}u_{yt}+\frac{\epsilon^2}{4}u_{xyt}^2}$; note that  $F\to f$ in the dispersionless limit $\epsilon \to 0$. The corresponding Euler-Lagrange equation is a sixth-order PDE for $u$,
$$
\left(\frac{L}{u_{xy}}\right)_{xy}+\left(\frac{L}{u_{xt}}\right)_{xt}+\left(\frac{L}{u_{yt}}\right)_{yt}-\frac{\epsilon}{2}\left( \ln \frac{2L-\epsilon u_{xyt}}{2L+\epsilon u_{xyt}}\right)_{xyt}=0.
$$
It was shown in \cite{XFP} that this PDE is an equivalent scalar form of the classical Darboux system,
\begin{equation*}
\partial _{k}\beta _{ij}=\beta _{ik}\beta _{kj},
\end{equation*}
governing rotation coefficients of triply-orthogonal coordinate systems in $\mathbb{R}^3$, see \cite{Darboux}.
Here $i\neq j\neq k \in \{1,2,3\}$ and $\partial_k$ denote partial derivatives with respect to the independent variables $x^k$. Introducing  potential $u$  via the relations $\partial_i\partial_ju=\beta_{ij}\beta_{ji}$ (that are compatible modulo Darboux system), one can rewrite the Darboux system as a single sixth-order equation for the potential $u$ which,  modulo simple rescalings,  coincides with the above PDE. It was shown in \cite{XFP} that this PDE is an alternative fully symmetric version of the `generating PDE of the KP hierarchy' as discussed in Nijhoff \cite{Nijhoff2024}.

\medskip

Integrable dispersive deformations of the Lagrangian densities from Cases $2-4$ are considerably more complicated: in contrast to Case 1, the corresponding dispersive Euler-Lagrange equations are no longer purely continuous,  constituting differential-difference equations with one, two and three discrete variables, respectively.
For example, in Case 2,  the Lagrangian density
$$
\begin{array}{c}
f=u_{xt}u_{yt}\sqrt{1-\frac{2u_{xy}}{u_{xt}u_{yt}}}-2u_{xy}\arctanh \sqrt{1-\frac{2u_{xy}}{u_{xt}u_{yt}}}
\end{array}
$$
possesses integrable differential-difference dispersive deformation with discrete variable $t$,
$$
\begin{array}{c}
F= \triangle_t u_x \triangle_t u_y  \tanh{\theta}-2 u_{xy} \, \theta+\epsilon\triangle_tu_{xy}\ln(1-\tanh{\theta}),
\end{array}
$$
where
\begin{equation*}
\begin{array}{c}
\tanh \theta=\frac{-\epsilon \triangle_t u_{xy}}{2 \triangle_t u_y \triangle_t u_x}+\sqrt{
\left(1-\frac{\epsilon \triangle_t u_{xy}}{2 \triangle_t u_y \triangle_t u_x}\right)^2-\frac{2u_{xy}}{ \triangle_t u_y \triangle_t u_x}}.
\end{array}
\end{equation*}
Here $\triangle_t=\frac{T_t-1}{\epsilon}$   is the discrete $t$-derivative  and $T_t$ denotes  $\epsilon$-shift in the variable $t$. Note that $F\to f$ in the dispersionless limit $\epsilon \to 0$. It was shown in \cite{XFP} that integrable dispersive deformations of the Lagrangian densities from Cases $2-4$ coincide with Lagrangian densities governing differential-difference and fully discrete versions of the Darboux system.

\section{Classification of integrable Lagrangians}
\label{sec:class}

\subsection{Quasilinear form of Euler-Lagrange equations}
\label{sec:quasi}

The Euler-Lagrange equation corresponding to Lagrangian (\ref{lag}) is a fourth-order PDE
for $u(x,y,t)$:
\begin{equation}
\left( \frac{\partial f}{\partial u_{xy}}\right) _{xy}+\left( \frac{\partial
f}{\partial u_{xt}}\right) _{xt}+\left( \frac{\partial f}{\partial u_{yt}}
\right) _{yt}=0.  \label{Euler}
\end{equation}
Setting  $u_{xy}=v_3, \ u_{xt}=v_2, \ u_{yt}=v_1$, one  can rewrite (\ref{Euler}) in
the form
\begin{equation}
(v_1)_x=(v_2)_y=(v_3)_t,\quad \left( f_{v_3}\right) _{xy}+\left(
f_{v_2}\right) _{xt}+\left( f_{v_1}\right) _{yt}=0.  \label{v}
\end{equation}
Introducing an auxiliary variable $p$ via the relations
\begin{equation}\label{pdef}
p_{y}=\left( f_{v_1}\right) _{y}+\left( f_{v_2}\right) _{x},\text{ \ }p_{t}=-\left( f_{v_3}\right)
_{x},
\end{equation}
one can rewrite (\ref{v}) as a first-order four-component conservative system,
\begin{equation}
\label{4}
(v_1)_x=(v_2)_y=(v_3)_t,\text{ \ } p_{y}=\left( f_{v_1}\right) _{y}+\left( f_{v_2}\right) _{x},\text{ \ }p_{t}=-\left( f_{v_3}\right)
_{x}.
\end{equation}
In matrix form,
\begin{equation*}
\mathbf {L w}_{x}+\mathbf {M w}_y+\mathbf{ N w}_t=0,
\end{equation*}
where $\mathbf {w}=(v_1,v_2,v_3,p)^{\text{T}}$ and $\mathbf{L},\mathbf{M}, \mathbf{N}$ are $4\times 4$ matrices:
\begin{equation*}
\mathbf{L}=\left(
\begin{array}{cccc}
1 & 0 & 0 & 0 \\
0 & 0 & 0 & 0 \\
f_{v_1v_2} & f_{v_2v_2} & f_{v_2v_3} & 0 \\
f_{v_1v_3} & f_{v_2v_3} & f_{v_3v_3} & 0
\end{array}
\right) ,\ \
\mathbf{M}=\left(
\begin{array}{cccc}
0 & -1 & 0 & 0 \\
0 & 1 & 0 & 0 \\
f_{v_1v_1} & f_{v_1v_2} & f_{v_1v_3} & -1 \\
0&0 & 0& 0
\end{array}
\right), \ \
\mathbf{N}=\left(
\begin{array}{cccc}
0 & 0 & 0 & 0 \\
0 & 0 & -1 & 0 \\
0 & 0 & 0 & 0 \\
0&0 & 0& 1
\end{array}
\right).
\end{equation*}
Quasilinear representation (\ref{4}) allows one to apply the integrability test based on the method of hydrodynamic reductions.

\subsection{Integrability conditions}
\label{sec:int}

The test of integrability based on the method of hydrodynamic reductions, see e.g. \cite{GT, FK04}, consists of seeking multiphase solutions of system (\ref{4}) in the form
\begin{equation}\label{uw}
v_i=v_i(R^1,R^2,\ldots,R^n), \quad p=p(R^1,R^2,\ldots,R^n),
\end{equation}
where the phases $R^i(x,y,t)$  satisfy a pair of commuting hydrodynamic-type systems:
\begin{equation}\label{R}
R^i_y = \mu^i(\mathbf{R})R^i_x, \quad
R^i_t = \lambda^i(\mathbf{R})R^i_x.
\end{equation}
The phases $R^i$ are known as Riemann invariants; note that their number, $n$, can be arbitrary. We recall that the commutativity conditions are equivalent to the following constraints for the characteristic speeds $\mu^i, \lambda^i$ \cite{Tsarev1, Tsarev2}:
\begin{equation}\label{comm}
\frac{\partial_j\mu^i}{\mu^j - \mu^i} = \frac{\partial_j\lambda^i}{\lambda^j - \lambda^i},
\end{equation}
$i\ne j, \ \partial_j=\partial_{R^j}$.  Substituting ansatz (\ref{uw}) into  (\ref{4}) and using (\ref{R}), (\ref{comm}), one obtains an overdetermined system  for  the unknowns $v_i, p, \mu^i, \lambda^i$, viewed as functions of $R^1, \dots, R^n$ (the so-called generalised Gibbons-Tsarev system, or GT-system).
System (\ref{4}) is said to be integrable by the method of hydrodynamic reductions if it possesses `sufficiently many' multi-phase solutions of  type (\ref{uw}), in other words, if the corresponding GT-system is involutive. Note that the coefficients of GT-system will depend on the density $f(v_1, v_2, v_3)$ and its partial derivatives. The requirement
that GT-system  is involutive imposes differential constraints for the Lagrangian density $f$, the so-called integrability conditions.

It was shown in  \cite{FK06} that the requirement of integrability by the method of hydrodynamic reductions implies a set of strong necessary conditions which, in many cases of interest, turn out to be also sufficient for integrability. In the present case, these necessary conditions are equivalent to the vanishing of  the Haantjes tensor of the matrix
$$
\mathbf{V}=(\mathbf{M}+\mu \mathbf{N})^{-1}(\mathbf{L}+\lambda \mathbf{N}),
$$
identically in the parameters $\lambda, \mu$. We recall that, given a matrix $\mathbf{V}(\mathbf{w})=(V^i_j(\mathbf{w}))$, its Haantjes tensor is defined by the formula
$$
H^i_{jk}=N^i_{sr}V^s_jV^r_k-N^s_{jr}V^i_sV^r_k-N^s_{rk}V^i_sV^r_j+N^s_{jk}V^i_rV^r_s,
$$
where $N$ is the Nijenhuis tensor:
$$
N^i_{jk}=V^s_j\partial_{w^s}V^i_k-V^s_k\partial_{w^s}V^i_j-V^i_s(\partial_{w^j}V^s_k-\partial_{w^k}V^s_j).
$$
Here we utilise the standard summation convention over the repeated indices. Calculating components of the Haantjes tensor $H$ and splitting them in $\lambda,\mu$, we obtain a number of third-order PDEs for the Lagrangian density $f$ that can be integrated once, giving  an involutive set of Monge-Amp\'ere equations:
\begin{equation}
\label{int}
\begin{array}{c}
f_{v_1v_1}f_{v_2v_2}-f_{v_1v_2}^2=a_3, \quad  f_{v_1v_1}f_{v_3v_3}-f_{v_1v_3}^2=a_2, \quad  f_{v_2v_2}f_{v_3v_3}-f_{v_2v_3}^2=a_1, \\
\ \\
f_{v_1v_2}f_{v_1v_3}-f_{v_1v_1}f_{v_2v_3}=p_1, \quad  f_{v_1v_2}f_{v_2v_3}-f_{v_2v_2}f_{v_1v_3}=p_2, \quad  f_{v_1v_3}f_{v_2v_3}-f_{v_3v_3}f_{v_1v_2}=p_3;
\end{array}
\end{equation}
here $a_i=const$ and each $p_i$ is a function of the argument $v_i$ only. Note that left-hand sides of  (\ref{int}) are $2\times 2$ minors of the Hessian matrix  $\mathbf{H}=Hess (f)$. Equations (\ref{int}) are explicitly solved in the next section, remarkably, the general case is quite nontrivial, expressed in terms of the Lobachevsky function.

\medskip

\noindent{\bf Theorem.}
{\it Conditions (\ref{int}) are  necessary and sufficient for integrability of system (\ref{4}) by the method of hydrodynamic reductions.}

\medskip

\noindent {\it Proof:} To show that conditions (\ref{int}) are indeed necessary and sufficient for integrability by the method of hydrodynamic reductions, let us look for multi-phase solutions of system (\ref{4}) in the form $v_i=v_i(R^{1},...,R^{n}), \ p=p(R^{1},...,R^{n})$, where the Riemann invariants $R^i$ satisfy  equations (\ref{R}).
Substituting this ansatz into  equations (\ref{4}), we obtain
\begin{equation}\label{v1i}
\partial_iv_1=\mu^{i}\partial_iv_2, ~~ \partial_iv_3=\frac{\mu^i}{\lambda^{i}}\partial_iv_2,
\end{equation}
and
\begin{align}
\begin{array}{c}
\mu^i\partial_ip=\left(f_{v_1v_1}{\mu^i}^2+2f_{v_1v_2}\mu^i+f_{v_1v_3}\frac{{\mu^i}^2}{\lambda^i}+f_{v_2v_2}+f_{v_2v_3}\frac{\mu^i}{\lambda^i}\right)\partial_iv_2, \\
{\lambda^i}\partial_i p=-\left(f_{v_1v_3}{\mu^i}+f_{v_2v_3}+f_{v_3v_3}\frac{\mu^i}{\lambda^i}\right)\partial_iv_2,
\end{array}
\label{V}
\end{align}
 where $\partial_i=\partial_{R^i}$ (no summation over $i$).
 Calculating consistency conditions for equations (\ref{v1i}),
 $$
 \partial_j\partial_iv_1=\partial_i\partial_jv_1, ~~
  \partial_j\partial_iv_3=\partial_i\partial_jv_3,
  $$
  and taking into account the commutativity conditions (\ref{comm}),
 we obtain
\begin{equation}
\begin{array}{c}
\partial _{i}\partial _{j}v_2=\frac{\partial _{j}\lambda ^{i}}{\lambda
^{j}-\lambda ^{i}}\partial _{i}v_2+\frac{\partial _{i}\lambda ^{j}}{\lambda
^{i}-\lambda ^{j}}\partial _{j}v_2, \\
\ \\
\lambda^j\partial_j\lambda^i\partial_iv_2+ \lambda^i\partial_i\lambda^j\partial_jv_2=0,
\end{array}
 \label{comp12}
\end{equation}
no summation.   Calculating consistency conditions for   equation  (\ref{V}), namely, calculating $\partial_i\partial_jp=\partial_j\partial_ip$ in two different ways using (\ref{comm}), (\ref{v1i}) and (\ref{comp12}), 
 we obtain  $\partial_j \lambda^i$ and $\partial_j \mu^i$ in the form
\begin{equation}
\partial_j \lambda^i=(\lambda^j-\lambda^i)\lambda^iB_{ij}\partial_jv_2, \ \ \ \ \
\partial_j \mu^i=(\mu^j-\mu^i)\lambda^iB_{ij}\partial_jv_2,
\label{lm}
\end{equation}
where  $B_{ij}$ are  symmetric rational expressions in $\lambda^i, \lambda^j, \mu^i, \mu^j$ whose coefficients depend on third-order partial derivatives of the function $f(v_1, v_2, v_3)$.
 Explicitly, one has
$$
B_{ij}=
\frac{f_{v_3v_3v_3}+f_{v_1v_1v_3}\lambda^j \lambda^i+f_{v_2v_2v_3} \rho^j \rho^i +f_{v_1v_3v_3} (\lambda^i+\lambda^j)+ f_{v_2v_3v_3}(\rho^i+\rho^j) + f_{v_1v_2v_3} (\rho^i \lambda^j+\rho^j \lambda^i) }
{(\lambda^i-\lambda^j) (\rho^i-\rho^j) f_{v_3v_3}}
$$
where
$\rho^i=\frac{\lambda^i}{\mu^{i}}$.
Finally, eliminating $\partial_ip$ from the two equations (\ref{V}), we obtain the dispersion relation,
\begin{equation}
f_{v_1v_1}{\lambda^i}^2+2f_{v_1v_2}\lambda^i{\rho^i}
+2f_{v_1v_3}{\lambda^i}+f_{v_2v_2}{\rho^i}^2+2f_{v_2v_3}{\rho^i}+f_{v_3v_3}=0,
\label{disp}
\end{equation}
which is a rational quartic in the $(\lambda, \mu)$-plane:
\begin{equation*}
f_{v_1v_1}{\lambda^i}^2{\mu^i}^2+2f_{v_1v_2}{\lambda^i}^2{\mu^i}+2f_{v_1v_3}{\lambda^i}{\mu^i}^2+f_{22}{\lambda^i}^2+2f_{v_2v_3}{\lambda^i}{\mu^i}+f_{v_3v_3}{\mu^i}^2=0.
\end{equation*}
 Applying the operator $\partial_j$ to the dispersion relation (\ref{disp}), using   (\ref{v1i}), \eqref{lm}, (\ref{disp})
and splitting the result  in $\lambda^i, \ \lambda^j, \ \rho^i, \ \rho^j$,
we obtain a system of nine third-order PDEs for $f$ that are equivalent to integrability conditions (\ref{int}).
Ultimately, hydrodynamic reductions of system (\ref{4}) are governed by the following system of Gibbons-Tsarev type:
\begin{equation}\label{GT}
\partial_j \lambda^i=(\lambda^j-\lambda^i)\lambda^iB_{ij}\partial_jv_2, \quad
\partial_j \mu^i=(\mu^j-\mu^i)\lambda^iB_{ij}\partial_jv_2, \quad \partial _{i}\partial _{j}v_2=(\lambda^i+\lambda^j)B_{ij}\partial _{i}v_2\partial _{j}v_2.
\end{equation}
It remains to verify by direct calculation that all compatibility conditions of the Gibbons-Tsarev system such as
$$
\partial_k\partial_j\lambda^i=\partial_j\partial_k\lambda^i, \quad \partial_k\partial_j\mu^i=\partial_j\partial_k\mu^i, \quad \partial_k\partial_j\partial_iv_2=\partial_j\partial_k\partial_iv_2,
$$
are satisfied identically modulo (\ref{int}) and (\ref{disp}). Thus, Gibbons-Tsarev system is involutive. This finishes the proof of integrability of system (\ref{4}) by the method of hydrodynamic reductions. $\square$

\subsection{Solving the integrability conditions: four  Lagrangian densities}
\label{sec:Lob}

In this section we solve equations (\ref{int}), obtaining four essentially different types of integrable Lagrangian densities $f$.
Note that equations (\ref{int}) allow one to represent  the inverse of the Hessian matrix $\mathbf{H}=Hess(f)$  in the form
\begin{equation}\label{F}
\left(
\begin{array}{ccc}
f_{v_1v_1}&f_{v_1v_2}&f_{v_1v_3}\\
f_{v_1v_2}&f_{v_2v_2}&f_{v_2v_3}\\
f_{v_1v_3}&f_{v_2v_3}&f_{v_3v_3}\\
\end{array}
\right)^{-1}=\frac{1}{\det \mathbf{H}}
\left(
\begin{array}{ccc}
a_1&p_3&p_2\\
p_3&a_2&p_1\\
p_2&p_1&a_3
\end{array}
\right).
\end{equation}
Taking the determinant of both sides we obtain
\begin{equation}\label{det}
(\det \mathbf{H})^2={a_1a_2a_3-a_1p_1^2-a_2p_2^2-a_3p_3^2+2p_1p_2p_3}.
\end{equation}
Inverting the matrix identity (\ref{F}) gives
\begin{equation}\label{F1}
\left(
\begin{array}{ccc}
f_{v_1v_1}&f_{v_1v_2}&f_{v_1v_3}\\
f_{v_1v_2}&f_{v_2v_2}&f_{v_2v_3}\\
f_{v_1v_3}&f_{v_2v_3}&f_{v_3v_3}\\
\end{array}
\right)=\frac{1}{\det \mathbf{H}}
\left(
\begin{array}{ccc}
a_2a_3-p_1^2&p_1p_2-a_3p_3&p_1p_3-a_2p_2\\
p_1p_2-a_3p_3&a_1a_3-p_2^2&p_2p_3-a_1p_1\\
p_1p_3-a_2p_2&p_2p_3-a_1p_1&a_1a_2-p_3^2
\end{array}
\right),
\end{equation}
where we use (\ref{det}) for $\det \mathbf{H}$. Consistency conditions of equations (\ref{F1}) lead to simple  ODEs for the functions $p_i(v_i)$:
$$
p_1'=\kappa(p_1^2-a_2a_3), \quad p_2'=\kappa(p_2^2-a_1a_3), \quad p_3'=\kappa(p_3^2-a_1a_2),
$$
where $\kappa$ is yet another arbitrary constant (which can be set equal to $-1$). The further analysis depends on how many constants among $a_i$ are equal to zero.
\medskip

\noindent{\bf Case 1. All constants  are zero.} In this case, without any loss of generality one can set $p_i=1/v_i$ which, up to a constant multiple, leads to the integrable Lagrangian density
$$
f=2 \sqrt{u_{xy}u_{xt}u_{yt}}.
$$

\medskip

\noindent{\bf Case 2. Two constants  are zero.} Then  one can also set $p_i=1/v_i$. Modulo (complex) rescalings this leads to the Lagrangian density (set $a_1=a_2=0, \, a_3=1$):
\begin{equation*}
\begin{array}{c}
f=u_{xt}u_{yt}\sqrt{\frac{2u_{xy}}{u_{xt}u_{yt}}-1}-2u_{xy}\arctan \sqrt{\frac{2u_{xy}}{u_{xt}u_{yt}}-1}.
\end{array}
\end{equation*}
It what follows, it will be more convenient to work with its hyperbolic version (set $p_i=-1/v_i,\ a_1=a_2=0,\ a_3=-1, \ \kappa=1$):
 \begin{equation*}
 \begin{array}{c}
 f=u_{xt}u_{yt}\sqrt{1-2\frac{u_{xy}}{u_{xt}u_{yt}}}-2u_{xy} \arctanh {\sqrt{1-2\frac{u_{xy}}{u_{xt}u_{yt}}}}.
 \end{array}
 \end{equation*}
\medskip
\noindent{\bf Case 3. One constant is zero.} This leads to the Lagrangian density (set $a_1=0,\ a_2=a_3=1, \ p_1=\coth v_1,\ p_2=1/v_2,\ p_3=1/v_3$):
\begin{equation*}
\begin{array}{c}
f=(u_{xt}-u_{xy})\arctan \frac{\sqrt{2u_{xt}u_{xy}\coth u_{yt}-u_{xt}^{2}-u_{xy}^{2}%
}}{u_{xt}-u_{xy}}-(u_{xt}+u_{xy})\arctan \frac{\sqrt{2u_{xt}u_{xy}\coth
u_{yt}-u_{xt}^{2}-u_{xy}^{2}}}{u_{xt}+u_{xy}}.
\end{array}
\end{equation*}
In what follows, it will be more convenient to work with its hyperbolic version  (set $a_1=0,\ a_2=a_3=-1, \ \kappa=1, \ p_1=-\coth v_1,\ p_2=-1/v_2,\ p_3=-1/v_3$):
\begin{equation*}
\begin{array}{c}
f=
 (u_{xt}-u_{xy}) \arctanh \frac{\sqrt{u_{xt}^{2}+u_{xy}^{2}- 2u_{xt}u_{xy}\coth u_{yt}}}{u_{xt}-u_{xy}}
-(u_{xt}+u_{xy}) \arctanh \frac{\sqrt{u_{xt}^{2}+u_{xy}^{2}- 2u_{xt}u_{xy}\coth u_{yt}}}{u_{xt}+u_{xy}}.
\end{array}
\end{equation*}

\medskip

\noindent{\bf Case 4. All constants  are nonzero.} This case  is most interesting and non-trivial, leading to remarkable formulas for Lagrangian densities $f$ that utilise the  Lobachevsky function and
spherical/hyperbolic laws of cosines. Below we discuss three geometrically different representations of a generic density from Case 4 (all equivalent modulo suitable complex rescalings).

\subsubsection{Lagrangian density corresponding to a spherical triangle}
\label{sec:striangle}

Setting $\kappa=-1, \ a_i=1$, we obtain $p_i'=1-p_i^2$ so that $p_i=\tanh v_i$. Equations (\ref{F1}) can be integrated once to yield
\begin{equation}\label{sphtr}
\begin{array}{c}
f_{v_1}=\arccos \left(\frac{\tanh v_1-\tanh v_2\tanh v_3}{\sech v_2\sech v_3}\right),
\\
f_{v_2}=\arccos \left(\frac{\tanh v_2 -\tanh v_1\tanh v_3}{\sech v_1\sech v_3}\right),
\\
f_{v_3}=\arccos \left( \frac{\tanh v_3-\tanh v_1\tanh v_2}{\sech v_1\sech v_2}\right),
\end{array}
\end{equation}
recall the notation $v_1=u_{yt}, \ v_2=u_{xt}, \ v_3=u_{xy}$. Equations (\ref{sphtr}) are consistent, however, the explicit reconstruction of $f$ is quite nontrivial, with an unexpected link to spherical trigonometry.  On the unit sphere $S^2$,  consider a spherical triangle  with interior angles $A, B, C$ and the opposite edge lengths $a, b, c$. The spherical laws of cosines are
\begin{equation}\label{sph}
\begin{array}{c}
\cos a=\cos b\cos c+\sin b\sin c\cos A,\\
\cos b=\cos a\cos c+\sin a\sin c\cos B,\\
\cos c=\cos a\cos b+\sin a\sin b\cos C,
\end{array}
\end{equation}
and
\begin{equation}\label{sph1}
\begin{array}{c}
\cos A=-\cos B\cos C+\sin B\sin C\cos a,\\
\cos B=-\cos A\cos C+\sin A\sin C\cos b,\\
\cos C=-\cos A\cos B+\sin A\sin B\cos c,
\end{array}
\end{equation}
respectively.  Note that the map $(A, B, C) \to (a, b, c)$ sending  angles of a spherical triangle  to its side lengths,  is integrable in the sense of multidimensional consistency \cite{PS} and is closely related to the discrete Darboux system \cite{BK, KS}.
Setting
\begin{equation} \label{pi}
\tanh v_1=\cos a,\text{ \ }\tanh v_2=\cos b,\text{ \ }\tanh v_3=\cos c
\end{equation}
and using (\ref{sph}), one can rewrite  (\ref{sphtr})  in the following  Schl\"afly-type form:
$$
df=A\,dv_1+B\,dv_2+C\,dv_3,
$$
equivalently,
\begin{equation}\label{S}
df=-\frac{A}{\sin a}da-\frac{B}{\sin b}db-\frac{C}{\sin c}dc.
\end{equation}
Recall that the classical Schl\"afly formula expresses the differential of the volume of a  polyhedron in terms of its side lengths and dihedral angles. Expression (\ref{S}), which can be viewed as a two-dimensional Schl\"afly formula, has appeared in \cite{Luo, Luo1} as a special  case of  one-parameter family of closed Schl\"afly-type forms associated with spherical triangles (case $h=0$ of Theorem 3.2(b) in \cite{Luo1}).  Note that the function $f$ defined by (\ref{S}) is essentially the `capacity'  of a spherical triangle which is defined as follows \cite{Luo2}. Let us view the unit sphere as the infinity of a hyperbolic 3-space (unit ball model). Given a spherical triangle, there are three circles bounding it.  The capacity of a spherical triangle is  the volume of the hyperbolic octahedron which is the convex hull of  six intersection points of the  circles  bounding the triangle.
 Expressions similar to (\ref{S}) have appeared before in the context of variational principles for circle packings on triangulated surfaces \cite{CdV, Bobenko, Bobenko1, Luo}.
Calculating the quadrature  (\ref{S}) is quite non-trivial, leading to a formula for $f$  in terms of the Lobachevsky function (see the Appendix):
\begin{equation}\label{fS}
\begin{array}{c}
f=Av_{1}+Bv_{2}+Cv_{3}+{\cal L}\left(\frac{2\pi-A-B-C}{2}\right) +{\cal L}\left(\frac{A+B-C}{2}\right) +{\cal L}\left(\frac{A+C-B}{2}\right) +{\cal L}\left(\frac{B+C-A}{2}\right).
\end{array}
\end{equation}
Note that $A, B, C$ are defined, as functions of $v_1, v_2, v_3$, via  relations (\ref{sph}), (\ref{pi}).

In terms of the side lengths $a, b, c$ and the angles $A, B, C$ of a spherical triangle,
 the Euler-Lagrange equation (\ref{v}) corresponding to the density $f$ takes the form
 \begin{equation}\label{ELhyptr}
 \begin{array}{c}
 A_{yt}+B_{xt}+C_{xy}=0,\\
 \ \\
 \frac{a_x}{\sin a}= \frac{b_y}{\sin b}= \frac{c_t}{\sin c},
 \end{array}
 \end{equation}
where we keep in mind the spherical cosine laws (\ref{sph}), (\ref{sph1}).

\subsubsection{Lagrangian density corresponding to a hyperbolic triangle}
\label{sec:htriangle}

 Setting $\kappa=-1, \ a_i=1$, we obtain $p_i'=1-p_i^2$, where we take $p_i=\coth v_i$. Equations (\ref{F1}) can be integrated once to yield
 the analogues of equations (\ref{sphtr}),
\begin{equation} \label{hyptr}
 \begin{array}{c}
f_{v_1}=\arccos\left(\frac{\coth{v_2}\coth{v_3}-\coth{v_1}}{{\rm csch}{v_2}{\rm csch}{v_3}}\right),
\\
f_{v_2}=\arccos\left(\frac{\coth{v_1}\coth{v_3}-\coth(v_2)}{{\rm csch}{v_1}{\rm csch}{v_3}}\right),
\\
f_{v_3}=\arccos\left(\frac{\coth{v_1}\coth{v_2}-\coth(v_3)}{{\rm csch}{v_1}{\rm csch}{v_2}}\right).
\end{array}
\end{equation}
In this case, the reconstruction of $f$ utilises  hyperbolic cosine laws: recall that for a hyperbolic triangle  with interior angles $A, B, C$ and the opposite edge lengths $a, b, c$, the hyperbolic laws of cosines are
\begin{equation}\label{hyp}
\begin{array}{c}
\cosh a=\cosh b\cosh c-\sinh b\sinh c\cos A,\\
\cosh b=\cosh a\cosh c-\sinh a\sinh c\cos B,\\
\cosh c=\cosh a\cosh b-\sinh a\sinh b\cos C,
\end{array}
\end{equation}
and
\begin{equation}\label{hyp1}
\begin{array}{c}
\cos A=-\cos B\cos C+\sin B\sin C\cosh a,\\
\cos B=-\cos A\cos C+\sin A\sin C\cosh b,\\
\cos C=-\cos A\cos B+\sin A\sin B\cosh c,
\end{array}
\end{equation}
respectively. Setting
\begin{equation} \label{pi1}
\coth v_1=\cosh a,\text{ \ }\coth v_2=\cosh b,\text{ \ }\coth v_3=\cosh c
\end{equation}
and using (\ref{hyp}), one can rewrite  (\ref{hyptr})  in the following  Schl\"afly-type form:
$$
df=A\,dv_1+B\,dv_2+C\,dv_3,
$$
equivalently,
\begin{equation}\label{H}
df=-\frac{A}{\sinh a}da-\frac{B}{\sinh b}db-\frac{C}{\sinh c}dc.
\end{equation}
Expression (\ref{H}) has appeared in (\cite{Luo1}, case $h=0$ of Theorem 3.2(c)), as a special  case of one-parameter family of closed Schl\"afly-type forms associated with hyperbolic triangles.
Calculating  quadrature  (\ref{H}) gives the following formula for $f$ (see the Appendix):
\begin{equation}\label{fH}
\begin{array}{c}
f={A}v_{1}+{B}v_{2}+{C}v_{3}-{\cal L}\left(\frac{{A}+{B}+{C}}{2}\right) +{\cal L}\left(\frac{{A}+{B}-{C}}{2}\right) +{\cal L}\left(\frac{{A}+{C}-{B}}{2}\right) +{\cal L}\left(\frac{{B}+{C}-{A}}{2}\right),
\end{array}
\end{equation}
where  $\cal L$  is the Lobachevsky function. Note that $A, B, C$ are defined, as functions of $v_1, v_2, v_3$, via  relations (\ref{hyp}), (\ref{pi1}).

In terms of the side lengths $a, b, c$ and the angles $A, B, C$ of a hyperbolic triangle,
 the Euler-Lagrange equation (\ref{v}) corresponding to the density $f$ takes the form
 \begin{equation}\label{ELhyptr}
 \begin{array}{c}
 A_{yt}+B_{xt}+C_{xy}=0,\\
 \ \\
 \frac{a_x}{\sinh a}= \frac{b_y}{\sinh b}= \frac{c_t}{\sinh c},
 \end{array}
 \end{equation}
where we keep in mind the hyperbolic cosine laws (\ref{hyp}), (\ref{hyp1}).

\subsubsection{Lagrangian density corresponding to a right-angled hyperbolic hexagon}
\label{sec:hexagon}

 Setting $\kappa=1, \ a_i=-1$, we obtain $p_i'=p_i^2-1$, where we take $p_i=-\coth v_i$. Equations (\ref{F1}) can be integrated once to yield
\begin{equation} \label{hyphex}
 \begin{array}{c}
f_{v_1}=\arccosh\left(\frac{\coth{v_2}\coth{v_3}-\coth{v_1}}{{\rm csch}{v_2}{\rm csch}{v_3}}\right),
\\
f_{v_2}=\arccosh\left(\frac{\coth{v_1}\coth{v_3}-\coth(v_2)}{{\rm csch}{v_1}{\rm csch}{v_3}}\right),
\\
f_{v_3}=\arccosh\left(\frac{\coth{v_1}\coth{v_2}-\coth(v_3)}{{\rm csch}{v_1}{\rm csch}{v_2}}\right).
\end{array}
\end{equation}
To reconstruct $f$, consider a convex right-angled hyperbolic hexagon with three non-adjacent edge lengths $L_1, L_2, L_3$ and  their opposite edge lengths $l_1, l_2, l_3$. The corresponding hyperbolic laws of cosines are (\cite{Beardon}, p. 160-161):
\begin{equation}\label{slhyphex}
\begin{array}{c}
\cosh{l_1}=-\cosh{l_2}\cosh{l_3}+\sinh{l_2}\sinh{l_3}\cosh L_1,\\
\cosh{l_2}=-\cosh{l_1}\cosh{l_3}+\sinh{l_1}\sinh{l_3}\cosh L_2,\\
\cosh{l_3}=-\cosh{l_1}\cosh{l_2}+\sinh{l_1}\sinh{l_2}\cosh L_3,
\end{array}
\end{equation}
and
\begin{equation}\label{s2hyphex}
\begin{array}{c}
\cosh L_1=-\cosh L_2\cosh L_3+\sinh L_2\sinh L_3\cosh{l_1},\\
\cosh L_2=-\cosh L_1\cosh L_3+\sinh L_1\sinh L_3\cosh{l_2},\\
\cosh L_3=-\cosh L_1\cosh L_2+\sinh L_1\sinh L_2\cosh{l_3},
\end{array}
\end{equation}
respectively. Setting
\begin{equation}\label{hex}
\coth v_i=-\cosh {l_i}
\end{equation}
and using (\ref{slhyphex}), one can rewrite  (\ref{hyphex})  in the following  Schl\"afly-type form:
$$
df=L_1\,dv_1+L_2\,dv_2+L_3\,dv_3,
$$
equivalently,
\begin{equation}\label{HH}
df=-\frac{L_1}{\sinh l_1}dl_1-\frac{L_2}{\sinh l_2}d l_2-\frac{L_3}{\sinh l_3}dl_3.
\end{equation}
Expression (\ref{HH}) has appeared  in (\cite{Luo1}), case $h=0$ of Theorem 3.2(d)), as a special  case of  one-parameter family of closed Schl\"afly-type forms associated with hyperbolic hexagons.  Calculating  quadrature  (\ref{HH}) gives the following formula for $f$ (see the Appendix):
\begin{equation}\label{fHH}
\begin{array}{c}
f={L_1}v_{1}+{L_2}v_{2}+{L_3}v_{3}\\
\ \\
-\phi\left(\frac{{L_1}+{L_2}+{L_3}}{2}\right) +\phi\left(\frac{{L_1}+{L_2}-{L_3}}{2}\right) +\phi\left(\frac{{L_1}+{L_3}-{L_2}}{2}\right) +\phi\left(\frac{{L_2}+{L_3}-{L_1}}{2}\right),
\end{array}
\end{equation}
where  $\phi(s) = - \int_0^s \ln  \cosh \xi \, d\xi$ is a hyperbolic analogue of the Lobachevsky function. Note that $L_i$ are defined,
 as functions of $v_i$, via relations (\ref{slhyphex}), ({\ref{hex}).


In terms of the side lengths $l_i, L_i$  of a hyperbolic hexagon,
 the Euler-Lagrange equation (\ref{v}) corresponding to the density $f$ takes the form
 \begin{equation}\label{ELhyphex}
 \begin{array}{c}
{(L_1)}_{yt}+{(L_2)}_{xt}+{(L_3)}_{xy}=0,\\
 \ \\
 \frac{(l_1)_x}{\sinh l_1}= \frac{(l_2)_y}{\sinh l_2}= \frac{(l_3)_t}{\sinh l_3},
 \end{array}
 \end{equation}
 where we keep in mind the hyperbolic cosine laws (\ref{slhyphex}), (\ref{s2hyphex}).

\section{Dispersionless Lax pairs}
\label{sec:dispLax}

In this section we construct dispersionless Lax pairs \cite{Zakharov} of systems (\ref{4}) corresponding to the Lagrangian densities $f$ from Cases 1-4 of section \ref{sec:Lob}. The structure of these Lax pairs clearly suggests that Cases 1-4 constitute commuting flows of the dispersionless KP hierarchy.
Dispersionless Lax pairs are sought in the form
$$
S_y=F(v_1, v_2, v_3,  p, S_x), \quad  S_t=G(v_1, v_2, v_3,  p, S_x),
$$
where the compatibility condition, $S_{xy}=S_{yx}$,  is required to be equivalent to system (\ref{4}). Direct calculation of the compatibility condition shows that $F$ and $G$ must be of the form
$$
F=F(v_3,\ g(S_x)+p), \quad G=G(v_2,\ g(S_x)+p-f_{v_1}),
$$
where the further specialisation of $F, G$ and $g$ depends of the particular form of the integrable Lagrangian density $f$.  Dispersionless Lax pairs play important role in the reconstruction of the corresponding integrable dispersive deformations \cite{Zakharov}, see Section \ref{sec:dispdef}.

\medskip
\noindent{\bf Case 1.} $f=2\sqrt{v_1v_2v_3}$. Dispersionless  Lax pair of the corresponding system  (\ref{4}) is
\begin{equation}\label{c1}
S_y=-\frac{v_3}{S_x+p}, \quad S_t=-\frac{v_2}{S_x+p-f_{v_1}},
\end{equation}
where $f_{v_1}= \frac{\sqrt{v_1v_2v_3}}{v_1}$.

\medskip
\noindent{\bf Case 2.}
 $
 f=v_1v_2\sqrt{2\frac{v_3}{v_1v_2}-1}-2v_3 \arctan {\sqrt{2\frac{v_3}{v_1v_2}-1}}.
 $
Dispersionless  Lax pair of the corresponding system  (\ref{4}) is
$$
S_y=-\frac{2v_3}{S_x+p}, \quad S_t=-2\arctan \frac{v_2}{S_x+p-f_{v_1}}
$$
where  $f_{v_1}=v_2\sqrt{2\frac{v_3}{v_1v_2}-1}$.
For our purposes, it will be more convenient to work with the hyperbolic version,
 \begin{equation}\label{f2hyp}
 f=v_1v_2\sqrt{1-2\frac{v_3}{v_1v_2}}-2v_3 \arctanh {\sqrt{1-2\frac{v_3}{v_1v_2}}},
 \end{equation}
with dispersionless Lax pair
\begin{equation}\label{c2}
S_y=\frac{2v_3}{S_x+p}, \quad S_t=2\arctanh \frac{v_2}{S_x+p-f_{v_1}},
\end{equation}
where $f_{v_1}=v_2\sqrt{1-2\frac{v_3}{v_1v_2}}$.

\medskip
\noindent{\bf Case 3.}
$
f=
 (v_{2}-v_{3})\arctan \frac{\sqrt{ 2v_{2}v_{3}\coth v_{1}-v_{2}^{2}-v_{3}^{2}}}{v_{2}-v_{3}}
-(v_{2}+v_{3})\arctan \frac{\sqrt{ 2v_{2}v_{3}\coth v_{1}-v_{2}^{2}-v_{3}^{2}}}{v_{2}+v_{3}}.
$
Dispersionless  Lax pair of the corresponding system  (\ref{4}) is
$$
S_y=-2 \arctan{\frac{v_3}{S_x+p}}, \quad S_t=-2 \arctan \frac{ v_2}{ S_x+p-f_{v_1}},
$$
where
$
f_{v_1}=\sqrt{ 2v_{2}v_{3}\coth v_{1}-v_{2}^{2}-v_{3}^{2}}.
$
It what follows, we will work with the  hyperbolic version,
\begin{equation}\label{f3hyp}
f=
 (v_{2}-v_{3}) \arctanh \frac{\sqrt{v_{2}^{2}+v_{3}^{2}- 2v_{2}v_{3}\coth v_{1}}}{v_{2}-v_{3}}
-(v_{2}+v_{3}) \arctanh \frac{\sqrt{v_{2}^{2}+v_{3}^{2}- 2v_{2}v_{3}\coth v_{1}}}{v_{2}+v_{3}},
\end{equation}
with dispersionless Lax pair
\begin{equation}\label{c3}
S_y=2 \arctanh \frac{v_3}{S_x+p}, \quad
S_t=2 \arctanh \frac{v_2}{S_x+p-f_{v_1}},
\end{equation}
where
$
f_{v_1}=\sqrt{v_{2}^{2}+v_{3}^{2}- 2v_{2}v_{3}\coth v_{1}}.
$

\medskip

\noindent{\bf Case 4.} We will consider all three cases separately:

\noindent   {\bf Lagrangian density corresponding to a spherical triangle} is defined by a quadrature (\ref{sphtr}):
\begin{equation*}
 \begin{array}{c}
f_{v_1}=\arccos\left(\frac{\tanh{v_1}-\tanh{v_2}\tanh{v_3}}{{\rm sech}{v_2}{\rm sech}{v_3}}\right),
\\
f_{v_2}=\arccos\left(\frac{\tanh{v_2}-\tanh{v_1}\tanh{v_3}}{{\rm sech}{v_1}{\rm sech}{v_3}}\right),
\\
f_{v_3}=\arccos\left(\frac{\tanh{v_3}-\tanh{v_1}\tanh{v_2}}{{\rm sech}{v_1}{\rm sech}{v_2}}\right).
\end{array}
\end{equation*}
Dispersionless  Lax pair of the corresponding system  (\ref{4}) is
 \begin{equation}\label{c4sptr}
S_y=\ln{\frac{e^{S_x+{\rm i}p+v_3}+1}{e^{S_x+{\rm i}p}-e^{v_3}}},
\quad
 S_t=\ln{\frac{e^{S_x+{\rm i}p-{\rm i}f_{v_1}+v_2}+1}{e^{S_x+{\rm i}p-{\rm i}f_{v_1}}-e^{v_2}}}.
\end{equation}

\noindent   {\bf Lagrangian density corresponding to a hyperbolic triangle} is defined by a quadrature (\ref{hyptr}):
\begin{equation*}
 \begin{array}{c}
f_{v_1}=\arccos\left(\frac{\coth{v_2}\coth{v_3}-\coth{v_1}}{{\rm csch}{v_2}{\rm csch}{v_3}}\right),
\\
f_{v_2}=\arccos\left(\frac{\coth{v_1}\coth{v_3}-\coth(v_2)}{{\rm csch}{v_1}{\rm csch}{v_3}}\right),
\\
f_{v_3}=\arccos\left(\frac{\coth{v_1}\coth{v_2}-\coth(v_3)}{{\rm csch}{v_1}{\rm csch}{v_2}}\right).
\end{array}
\end{equation*}
Dispersionless  Lax pair of the corresponding system  (\ref{4}) is
 \begin{equation}\label{c4hyptr}
 S_y=\ln{\frac{e^{S_x+{\rm i}p+v_3}-1}{e^{S_x+{\rm i}p}-e^{v_3}}},
\quad
 S_t=\ln{\frac{e^{S_x+{\rm i}p-{\rm i}f_{v_1}+v_2}-1}{e^{S_x+{\rm i}p-{\rm i}f_{v_1}}-e^{v_2}}}.
\end{equation}
Note that dispersionless Lax pair (\ref{c4sptr}) goes to (\ref{c4hyptr}) under the transformation
$$
v_k\to v_k+\frac{\pi {\rm i}}{2}, \quad f\to f, \quad p\to p,\quad S\to S+\frac{\pi {\rm i}}{2}(x+y+t).
$$

\noindent   {\bf Lagrangian density corresponding to a hyperbolic hexagon} is defined by a quadrature (\ref{hyphex}):
\begin{equation*}
 \begin{array}{c}
f_{v_1}=\arccosh\left(\frac{\coth{v_2}\coth{v_3}-\coth{v_1}}{{\rm csch}{v_2}{\rm csch}{v_3}}\right),
\\
f_{v_2}=\arccosh\left(\frac{\coth{v_1}\coth{v_3}-\coth(v_2)}{{\rm csch}{v_1}{\rm csch}{v_3}}\right),
\\
f_{v_3}=\arccosh\left(\frac{\coth{v_1}\coth{v_2}-\coth(v_3)}{{\rm csch}{v_1}{\rm csch}{v_2}}\right).
\end{array}
\end{equation*}
Dispersionless  Lax pair of the corresponding system  (\ref{4}) is
\begin{equation}\label{c4hyp}
S_y=\ln{\frac{e^{S_x+p+v_3}-1}{e^{S_x+p}-e^{v_3}}}, \quad  S_t=\ln{\frac{e^{S_x+p-f_{v_1}+v_2}-1}{e^{S_x+p-f_{v_1}}-e^{v_2}}}.
\end{equation}
Note that dispersionless Lax pair (\ref{c4sptr}) goes to (\ref{c4hyp}) under the transformation
$$
v_k\to v_k+\frac{\pi {\rm i}}{2}, \quad f\to -{\rm i}f, \quad p\to -{\rm i} p,\quad S\to S+\frac{\pi {\rm i}}{2}(x+y+t).
$$

\section{Dispersive deformations}
\label{sec:dispdef}

In this section, we construct integrable dispersive deformations of dispersionless Lax pairs from the four cases of Section \ref{sec:dispLax}.
Remarkably, Case 1 has a purely continuous dispersive deformations, whereas Cases 2-4 lead to differential-difference equations with one, two and three discrete variables, respectively. These dispersive deformations are shown to coincide with scalar Lagrangian formulations of the classical Darboux system including its differential, semi-discrete and fully discrete versions \cite{XFP}.
We denote by $T_x, \, T_y, \,T_t$   the $\epsilon$-shifts in the corresponding variables, and by $\triangle_x=\frac{T_x-1}{\epsilon}, \, \triangle_y=\frac{T_y-1}{\epsilon}, \, \triangle_t=\frac{T_t-1}{\epsilon}$ the corresponding discrete derivatives.

\subsection{Case 1.}
\label{sec:Case1}

Here we use $f=2\sqrt{u_{xy}u_{xt}u_{yt}}$. The corresponding dispersionless Lax pair has the general form
\begin{equation}\label{lax1}
S_y=\frac{\alpha}{S_x+m}, \quad S_t=\frac{\beta}{S_x+n}.
\end{equation}
The compatibility condition $S_{yt}=S_{ty}$ gives the  dispersionless system,
\begin{equation}\label{Case1}
\begin{array}{c}
m_t=\left(\frac{\beta}{m-n}\right)_x, \quad  n_y=\left(\frac{\alpha}{n-m}\right)_x,\\
\ \\
\alpha_t=\beta_y=-\left(\frac{\alpha\beta}{(m-n)^2}\right)_x.
\end{array}
\end{equation}
As noted in (\cite{Zakharov}, formula (9)), system (\ref{Case1}) comes from the Lagrangian
\begin{equation}\label{ZL}
\int \left(\alpha\, \partial_x^{-1}m_t-{\beta}\, \partial_x^{-1}n_y-\frac{\alpha\beta}{m-n}\right)\, dxdydt.
\end{equation}
Under the substitution $\alpha=-v_3, \ \beta=-v_2, \ m=p, \ n=p-f_{v_1}$ where
$f_{v_1}= {\frac{\sqrt{v_1v_2v_3}}{v_1}}$,
dispersionless Lax pair (\ref{lax1}) goes to (\ref{c1}), the last two equations (\ref{Case1}) will be satisfied identically, while the first two equations (\ref{Case1}) lead to equations (\ref{pdef}) that are equivalent to the Euler-Lagrange equation for the Lagrangian density $f=2\sqrt {v_1v_2v_3}$ (recall that $v_1=u_{yt}, \ v_2=u_{xt}, \ v_3=u_{xy}$).

Dispersive deformation of dispersionless Lax equations (\ref{lax1}) is
\begin{equation}\label{Lax1}
\begin{array}{c}
\epsilon^2\psi_{xy}+\epsilon m \psi_y=\alpha \psi, \quad \epsilon^2\psi_{xt}+\epsilon n \psi_t=\beta\psi,\\
\ \\
\epsilon^2\psi_{yt}+\frac{\epsilon^2m_t+\epsilon \beta}{m-n}\ \psi_y+\frac{\epsilon^2n_y+\epsilon \alpha}{n-m}\ \psi_t+\epsilon\frac{\beta_y-\alpha_t}{m-n}\ \psi=0,
\end{array}
\end{equation}
where the last equation follows from the consistency condition of the first two equations (\ref{Lax1}), namely, from $(\psi_{xy})_t=(\psi_{xt})_y$.
The compatibility conditions of dispersive Lax equations (\ref{Lax1}), obtained by calculating $\psi_{xyt}$ from the three equations (\ref{Lax1}), lead to an integrable dispersive deformation of system (\ref{Case1}),
\begin{equation}\label{Case1disp1}
\begin{array}{c}
m_t=\left(\frac{\beta+\epsilon m_t}{m-n}\right)_x, \quad  n_y=\left(\frac{\alpha+\epsilon n_y}{n-m}\right)_x,\\
\ \\
\alpha_t=\beta_y=-\left(\frac{(\alpha+\epsilon n_y)(\beta+\epsilon m_t)}{(m-n)^2}\right)_x.
\end{array}
\end{equation}
Note that dispersionless Lax pair (\ref{lax1}) follows from  dispersive Lax equations (\ref{Lax1}) via the standard substitution  $\psi=e^{S/\epsilon}$ \cite{Zakharov}, followed by cancelling the common factor  $e^{S/\epsilon}$ and taking the limit $\epsilon\to 0$. Here the first two Lax equations (\ref{Lax1}) give  dispersionless equations (\ref{lax1}), while the third equation (\ref{Lax1}) gives no additional constraints for $S$; note that the $\psi$-term in the third equation (\ref{Lax1}) can be set equal to zero (as $\alpha_t=\beta_y$).
Dispersive deformation of Lagrangian (\ref{ZL}) is
\begin{equation}\label{ZL1disp}
\int \left((\alpha+\frac{\epsilon}{2}n_y)\, \partial_x^{-1}m_t-(\beta+\frac{\epsilon}{2}m_t)\, \partial_x^{-1}n_y-\frac{(\alpha+\epsilon n_y)(\beta+\epsilon m_t)}{m-n}\right)\, dxdydt.
\end{equation}
The last two equations of system (\ref{Case1disp1}) suggest the introduction of a potential $u$ such that
\begin{equation}\label{potu1}
\alpha=-u_{xy}, \quad \beta=-u_{xt}, \quad \frac{(\alpha+\epsilon n_y)(\beta+\epsilon m_t)}{(m-n)^2}=u_{yt}.
\end{equation}
Our next goal is to rewrite system (\ref{Case1disp1}) in terms of the potential $u$. In what follows, we will use the notation $L=\sqrt{ u_{xy}u_{xt}u_{yt}+\frac{\epsilon^2}{4}u_{xyt}^2}.$

\begin{proposition} \label{pc1} System (\ref{Case1disp1}) can be written as a single sixth-order PDE for the potential $u$,
\begin{equation}\label{genKP}
\left(\frac{L}{u_{xy}}\right)_{xy}+\left(\frac{L}{u_{xt}}\right)_{xt}+\left(\frac{L}{u_{yt}}\right)_{yt}- \left(\frac{\epsilon}{2} \ln \frac{L-\frac{\epsilon}{2} u_{xyt}}{L+ \frac{\epsilon}{2}  u_{xyt}}\right)_{xyt}=0.
\end{equation}
Equation (\ref{genKP}) is represented in Euler-Lagrange form corresponding to a  Lagrangian $\int F\, dxdydt$, with the Lagrangian density
\begin{equation}\label{f1disp}
F=2 L+\frac{\epsilon}{2}  u_{xyt} \ln \frac{L-\frac{\epsilon}{2}  u_{xyt}}{L+\frac{\epsilon}{2}  u_{xyt}}.
\end{equation}
In the dispersionless limit $\epsilon \to 0$,  dispersive density $F$ reduces to the dispersionless density $f=2\sqrt{u_{xy}u_{xt}u_{yt}}$.
\end{proposition}

\noindent {\it Proof:} Parametrising the last relation (\ref{potu1}) in the form
$$
\begin{array}{c}
\frac{\alpha+\epsilon n_y}{n-m}= e^{\varphi}, \quad \frac{\beta+\epsilon m_t}{m-n}=-u_{yt}\, e^{-\varphi},
\end{array}
$$
and substituting into the first two equations (\ref{Case1disp1}), we obtain two relations that are equivalent to
\begin{equation}\label{c1short}
u_{xt}e^{2\varphi}+\epsilon{ u_{xyt}}e^{\varphi} -{u_{xy}}{u_{yt}}=0, \quad n-m+ e^{-\varphi} {u_{xy}}-\epsilon \varphi_x=0.
\end{equation}
The first relation (\ref{c1short}) can be solved for  $\varphi$; the solution can be represented in either of the equivalent forms,
$$
e^{\varphi}=\frac{L-\frac{\epsilon}{2}  u_{xyt}}{u_{xt}}, \quad e^{-\varphi}=\frac{L+\frac{\epsilon}{2} u_{xyt}}{ u_{xy}u_{yt}},\quad
\varphi=\frac{1}{2}\ln \frac{(L-\frac{\epsilon}{2}  u_{xyt})u_{xy}u_{yt}}{(L+\frac{\epsilon}{2}  u_{xyt})u_{xt}},
$$
so that the first two equations (\ref{Case1disp1}) reduce to
\begin{equation}\label{sub1}
m_t=\left(-\frac{L+\frac{\epsilon}{2}  u_{xyt}}{ u_{xy}}\right)_x,\quad n_y=\left(\frac{L-\frac{\epsilon}{2} u_{xyt}}{ u_{xt}}\right)_x,
\end{equation}
respectively. Applying to the second relation (\ref{c1short}) the operator $\partial_y\partial_t$ and using (\ref{sub1})  leads, on simplification, to the sixth-order PDE (\ref{genKP}) for $u$.
To reconstruct the corresponding Lagrangian, note that for a Lagrangian density of the form $F=F(u_{xy},u_{xt}, u_{yt}, u_{xyt})$, the Euler-Lagrange equation is
$$
\left(\frac{\partial F}{\partial u_{xy}}\right)_{xy}+\left(\frac{\partial F}{\partial u_{xt}}\right)_{xt}+\left(\frac{\partial F}{\partial u_{yt}}\right)_{yt}-\left(\frac{\partial F}{\partial u_{xyt}}\right)_{xyt}=0.
$$
Comparison with (\ref{genKP}) gives all first-order derivatives of $F$:
$$
\frac{\partial F}{\partial u_{xy}}=\frac{L}{u_{xy}}, \quad \frac{\partial F}{\partial u_{xt}}=\frac{L}{u_{xt}}, \quad\frac{\partial F}{\partial u_{yt}}=\frac{L}{u_{yt}}, \quad\frac{\partial F}{\partial u_{xyt}}=\frac{\epsilon}{2} \ln \frac{L-\frac{\epsilon}{2}  u_{xyt}}{L+\frac{\epsilon}{2} u_{xyt}}.
$$
This system for $F$ is consistent and, on integration, leads to  Lagrangian density (\ref{f1disp}).
It was demonstrated in \cite{XFP} that equation (\ref{genKP}) constitutes an alternative scalar form of the classical Darboux system and is equivalent to the `generating PDE of the KP hierarchy' as discussed in \cite{Nijhoff2024}.  $\square$

\medskip
{\bf \noindent Remark 3.}
It would be natural to expect that  Lagrangian density (\ref{f1disp}) can be obtained from the Lagrangian density of  (\ref{ZL1disp}). One should note however that modifying Lagrangian (\ref{ZL1disp}) using the corresponding Euler-Lagrange equations (\ref{Case1disp1}) is not an allowed operation and does not necessarily produce an equivalent Lagrangian. A rigorous way of obtaining  (\ref{f1disp}) directly  from (\ref{ZL1disp}) is yet to be worked out.

\subsection{Case 2}
\label{sec:Case2}

Here we use formula (\ref{f2hyp}) for $f$. The corresponding dispersionless Lax pair has the general form
\begin{equation}\label{lax2}
S_y=\frac{\alpha}{S_x+m}, \quad S_t=2\arctanh \frac{\beta}{S_x+n}.
\end{equation}
The compatibility condition $S_{yt}=S_{ty}$ gives the  dispersionless equations,
\begin{equation}\label{Case2}
\begin{array}{c}
m_t=\left(\ln \frac{\beta+m-n}{\beta+n-m}\right)_x, \quad
n_y=\left( \frac{\alpha (m-n)}{(\beta+m-n)(\beta+n-m)}\right)_x,\\
\ \\
\alpha_t=2\beta_y=\left( \frac{2\alpha\beta}{(\beta+m-n)(\beta+n-m)}\right)_x,
\end{array}
\end{equation}
which come from the Lagrangian
\begin{equation}\label{ZL2}
\int \left(\alpha \partial_x^{-1}m_t-2\beta\, \partial_x^{-1}n_y-\alpha \ln \frac{\beta+m-n}{\beta+n-m}\right)\, dxdydt.
\end{equation}
Under the substitution $\alpha=2v_3,\, \beta=v_2,\, m=p, \, n=p-f_{v_1}$  where
$f_{v_1}=v_2\sqrt{1-2\frac{v_3}{v_1v_2}},$
 dispersionless Lax pair (\ref{lax2}) goes to (\ref{c2}), the last two equations (\ref{Case2}) will be satisfied identically,  while the first two equations (\ref{Case2}) lead to equations (\ref{pdef}) that are equivalent to the Euler-Lagrange equation for the corresponding Lagrangian density $f$.

To construct dispersive deformation of dispersionless Lax pair (\ref{lax2}), we first rewrite it in equivalent form,
\[
S_yS_x+mS_y=\alpha,\quad (e^{S_t}-1)(S_x+n-{\beta})=2\beta.
\]
Its dispersive deformation  is
\begin{equation}\label{Lax2}
\begin{array}{c}
\epsilon^2 \psi_{xy}+\epsilon m \psi_y=\alpha \psi,\quad
 \epsilon^2 \triangle_t\psi_x+\epsilon(n-{\beta})\triangle_t\psi=2\beta\psi,
\\
\epsilon^2 (\beta-n+T_t m)\triangle_t{\psi}_y+\epsilon (2\beta+\epsilon \triangle_t m )\psi_y =\epsilon [\alpha+\epsilon (\triangle_t \alpha  -{\beta}_y+n_y)]\triangle_t \psi+\epsilon (\triangle _t \alpha-2\beta_y)\psi;
\end{array}
\end{equation}
note that the last equation (\ref{Lax2}) follows
from the consistency condition of the first two.
The compatibility conditions of dispersive Lax equations (\ref{Lax2})
lead to an integrable dispersive deformation
of system (\ref{Case2}),
\begin{equation}\label{Case2disp0}
\begin{array}{c}
\triangle_t{\alpha}=2 \beta_y,\quad
\triangle_t{m}=\left(\ln\frac{\beta-n+T_t{m}}{\beta+n-m}\right)_x,\quad
\triangle_t{(\alpha m)}=2 (\beta n)_y,\\
(\beta+n)_y=\left(\frac{\alpha+\epsilon(\beta+n)_y}{\beta+n-m}\right)_x,
\end{array}
\end{equation}
which is equivalent to
\begin{equation}\label{Case2disp}
\begin{array}{c}
\triangle_t{m}=\left(\ln\frac{\beta-n+T_t{m}}{\beta+n-m}\right)_x,\quad
n_y=\left(\frac{(\alpha+\epsilon(\beta+n)_y)(T_t{m}+m-2n)}{2(\beta-n+T_t{m})(\beta+n-m)}\right)_x,
\\
\triangle_t{\alpha}=2\beta_y=\left(\frac{(\alpha+\epsilon(\beta+n)_y)(2\beta+\epsilon \triangle_t{m} )}{(\beta-n+T_t{m})(\beta+n-m)}\right)_x.
\end{array}
\end{equation}
Dispersionless Lax pair (\ref{lax2}) follows from  dispersive Lax equations  (\ref{Lax2}) via the standard substitution  $\psi=e^{S/\epsilon}$,
followed by cancelling the common factor  $e^{S/\epsilon}$ and taking the limit $\epsilon\to 0$. Here the first two Lax equations (\ref{Lax2}) give  dispersionless equations (\ref{lax2}), while the third equation (\ref{Lax2}) gives no additional constraints for $S$; note that the $\psi$-term in the third equation (\ref{Lax2}) can be set equal to zero.
The corresponding dispersive deformation of Lagrangian (\ref{ZL2}) is
\begin{equation}\label{ZL2disp}
\int \left((\alpha+\epsilon(\beta+n)_y)\, \partial_x^{-1}\triangle_t m-2\beta\, \partial_x^{-1}n_y-(\alpha+\epsilon(\beta+n)_y) \ln\frac{\beta-n+T_t{m}}{\beta+n-m}\right)\, dxdy\delta t,
\end{equation}
where integration over the discrete variable $t$, indicated by $\delta t$, is understood as  summation over all $T_t$-translates of the Lagrangian density (\ref{ZL2disp}). The last two equations of system (\ref{Case2disp}) suggest the introduction of a potential $u$ such that
\begin{equation}\label{potu2}
\alpha=2 u_{xy}, \quad \beta=\triangle_t u_{x}, \quad \frac{(\alpha+\epsilon(\beta+n)_y)(2\beta+\epsilon \triangle_t{m} )}{(\beta-n+T_t{m})(\beta+n-m)} =2\triangle_t u_{y}.
\end{equation}
Our next goal is to rewrite system (\ref{Case2disp}) in terms of the potential $u$. In what follows, we will use the quantity $\theta$ defined as
\begin{equation}\label{theta2}
\tanh \theta=\frac{-\epsilon \triangle_t u_{xy}}{2 \triangle_t u_y \triangle_t u_x}+\sqrt{
\left(1-\frac{\epsilon \triangle_t u_{xy}}{2 \triangle_t u_y \triangle_t u_x}\right)^2-\frac{2u_{xy}}{ \triangle_t u_y \triangle_t u_x}}.
\end{equation}

\begin{proposition} \label{pc2} System (\ref{Case2disp}) can be written as a single differential-difference equation for the potential  $u$,
\begin{equation}\label{C2D}
(-2\theta)_{xy}+\triangle_t\big(\triangle_t u_y \tanh{\theta})_x
+\triangle_t(\triangle_t u_x \tanh{\theta})_y-\epsilon \triangle_t(\theta-\ln \cosh \theta)_{xy}=0.
\end{equation}
Equation (\ref{C2D}) is represented in Euler-Lagrange form corresponding to a  Lagrangian $\int F\, dxdy\delta t$, with the Lagrangian density
\begin{equation}\label{f2disp}
\begin{array}{c}
F= \triangle_t u_x \triangle_t u_y  \tanh{\theta}-2 u_{xy} \, \theta+\epsilon\triangle_tu_{xy}\ln(1-\tanh{\theta}).
\end{array}
\end{equation}
In the dispersionless limit $\epsilon \to 0$, we have $\tanh \theta=\sqrt{1-\frac{2u_{xy}}{ u_{xt} u_{yt}}}$, and the dispersive Lagrangian density $F$ reduces to the dispersionless density $f$ given by (\ref{f2hyp}).
\end{proposition}

\noindent {\it Proof:}  Setting $\frac{\beta-n+T_t{m}}{\beta+n-m}=e^{2\varphi}$, the first equation of system (\ref{Case2disp}) implies the relations
\begin{equation}\label{Case2disp2}
\begin{array}{c}
n-m+\triangle_t u_x \tanh{\varphi}-\epsilon \varphi_x (1- \tanh{\varphi})=0, \quad  \triangle_t m=2\varphi_x.
\end{array}
\end{equation}
Combining the last relation (\ref{potu2}) with the second equation (\ref{Case2disp})  implies the equation $n_y=(\triangle_t u_y \tanh{\varphi})_x$, as well as a quadratic equation for $\tanh \varphi$,
\[
 \tanh^2{\varphi}+\frac{\epsilon \triangle_t u_{xy}}{
  \triangle_t u_y \triangle_t u_x} \tanh{\varphi}+\frac{2u_{xy}+\epsilon \triangle_t u_{xy}}{ \triangle_t u_y \triangle_t u_x}-1=0,
\]
with one of its solutions given by (\ref{theta2}). Thus, we can set $\varphi=\theta$. Ultimately, we obtain the relations
\begin{equation}\label{rel2}
n-m+\triangle_t u_x \tanh{\theta}-\epsilon \theta_x (1- \tanh{\theta})=0, \quad  \triangle_t m=2\theta_x, \quad n_y=(\triangle_t u_y \tanh{\theta})_x.
\end{equation}
To eliminate $m$ and $n$, we apply the operator $\partial_y\triangle_t$ to the first relation (\ref{rel2}), which results in equation (\ref{C2D}).
To reconstruct the corresponding Lagrangian, note that for a Lagrangian density  of the form $F=F(u_{xy},\, \triangle_tu_x,\, \triangle_tu_y,\, \triangle_tu_{xy})$, the Euler-Lagrange equation is
$$
\left(\frac{\partial F}{\partial u_{xy}}\right)_{xy}+\triangle_t\left(\frac{\partial F}{\partial (\triangle_tu_{x})}\right)_{x}+\triangle_t\left(\frac{\partial F}{\partial (\triangle_tu_{y})}\right)_{y}-\triangle_t\left(\frac{\partial F}{\partial (\triangle_tu_{xy})}-\epsilon\frac{\partial F}{\partial u_{xy}}\right)_{xy}=0.
$$
Comparison with (\ref{C2D}) gives  all first-order derivatives of $F$:
$$
\frac{\partial F}{\partial u_{xy}}=-2\theta, \quad \frac{\partial F}{\partial (\triangle_tu_{x})}=\triangle_t u_y \tanh{\theta}, \quad \frac{\partial F}{\partial (\triangle_tu_{y})}=\triangle_t u_x \tanh{\theta}, \quad \frac{\partial F}{\partial (\triangle_tu_{xy})}=-\epsilon(\theta+\ln \cosh \theta).
$$
This system for $F$ is consistent and, on integration,  leads to the Lagrangian density (\ref{f2disp}).
It was demonstrated in \cite{XFP} that equation (\ref{C2D}) constitutes an alternative scalar form of a differential-difference version of the classical Darboux system with one discrete variable.
Indeed, introducing $L, \  q$ such that
$
\tanh \theta=\frac{-\epsilon \triangle_t u_{xy}-L}{2 \triangle_tu_{x}\triangle_t u_y}, \ q=L+2\triangle_tu_{x}\triangle_tu_{y}-\epsilon  \triangle_t u_{xy},
$
one has
\[
L^2= {(2 \triangle_tu_{x}\triangle_t u_y-\epsilon \triangle_tu_{xy})^2- 8u_{xy}\triangle_tu_{x}\triangle_t u_y },
\]
\[
\tanh{\theta}=\frac{4 u_{xy}}{q}-1,\quad  \theta=-\frac{1}{2}\ln \left(\frac{q}{2 u_{xy}}-1\right).
\]
Thus, the Lagrangian density
\eqref{f2disp} can be written as
\begin{equation*}\label{f2disp2}
\begin{array}{c}
F
=-\frac{\epsilon }{2} \triangle_t u_{xy}-\frac{L}{2}+u_{xy}\ln\left(\frac{q}{2 u_{xy}}-1\right)+\epsilon \triangle_t u_{xy} \ln\left(2-\frac{4 u_{xy}}{q}\right).
\end{array}
\end{equation*}
Modulo total derivatives and rescalings, it is equivalent to the one constructed in \cite{XFP} for the semi-discrete Darboux system with one discrete variable.
$\square$

\subsection{Case 3}
\label{sec:Case3}

Here we use formula (\ref{f3hyp}) for $f$. The corresponding dispersionless Lax pair has the general form
\begin{equation}\label{lax3}
S_y=2\arctanh \frac{\alpha}{S_x+m}, \quad S_t=2\arctanh \frac{\beta}{S_x+n}.
\end{equation}
The compatibility condition $S_{yt}=S_{ty}$ gives the  dispersionless equations,
\begin{equation}\label{Case3}
\begin{array}{c}
m_t=\frac{1}{2}\left(\ln \frac{(\alpha+\beta +m-n)(a-\beta+n-m)}{(\alpha+\beta +n-m)(\alpha-\beta+m-n)}\right)_x, \quad
n_y=\frac{1}{2}\left(\ln \frac{(\alpha+\beta +n-m)(a-\beta+n-m)}{(\alpha+\beta +m-n)(\alpha-\beta+m-n)}\right)_x,\\
\ \\
\alpha_t=\beta_y=\frac{1}{2}\left(\ln \frac{(\alpha+\beta +m-n)(\alpha+\beta +n-m)}{(\alpha-\beta+m-n)(\alpha-\beta+n-m)}\right)_x.
\end{array}
\end{equation}
System (\ref{Case3}) comes from the Lagrangian
\begin{equation}\label{ZL3}
\int \left(\alpha\, \partial_x^{-1}m_t-{\beta}\, \partial_x^{-1}n_y-l\right)\, dxdydt
\end{equation}
where
$$
l=\varphi(\alpha+\beta +m-n)+\varphi(\alpha-\beta+n-m)-\varphi(\alpha-\beta+m-n)-\varphi(\alpha+\beta +n-m),
$$
$\varphi(s)=\frac{1}{2}s\ln s$.
Under the substitution $\alpha=v_3, \, \beta=v_2, \, m=p, \, n=p-f_{v_1}$ where $f_{v_1}=\sqrt{v_{2}^{2}+v_{3}^{2}- 2v_{2}v_{3}\coth v_{1}}$, dispersionless Lax pair (\ref{lax3}) goes to (\ref{c3}), the last two equations (\ref{Case3}) will be satisfied identically, while the first two equations (\ref{Case3})  lead to equations (\ref{pdef}) that are equivalent to the Euler-Lagrange equation for the corresponding Lagrangian density $f$.


To construct dispersive deformation of dispersionless Lax pair (\ref{lax3}), we first rewrite it in equivalent form,
\begin{equation*}
 (e^{S_t}-1)(S_x+n-{\beta})=2\beta, \quad  (e^{S_y}-1)(S_x+m-\alpha)=2\alpha.
\end{equation*}
Its dispersive deformation  is
\begin{equation}\label{Lax3}
\begin{array}{c}
 \epsilon^2 \triangle_t\psi_x+\epsilon(n-{\beta})\triangle_t\psi=2\beta\psi,\quad
  \epsilon^2 \triangle_y\psi_x+\epsilon(m-\alpha)\triangle_y\psi=2\alpha\psi,
\\
\epsilon \triangle_t\left[\epsilon(\alpha-m)\triangle_y \psi +2\alpha\psi\right]=\epsilon \triangle_y\left[\epsilon(\beta-n)\triangle_t \psi +2\beta\psi\right];
\end{array}
\end{equation}
note that the last equation (\ref{Lax3}) follows
from the consistency condition of the first two.
The compatibility conditions of dispersive Lax equations (\ref{Lax3})
lead to an integrable dispersive deformation
of system (\ref{Case3}),
\begin{equation*}
\begin{array}{c}
\triangle_t \alpha=\triangle_y \beta,\qquad\qquad \triangle_t (\alpha m)=\triangle_y (\beta n),\\
\triangle_t (\alpha+m)=\left(\ln \frac{T_t \alpha+\beta +T_t m-n}{\alpha-\beta+m-n}\right)_x,\quad
\triangle_y (\beta+n)=\left(\ln \frac{T_t \alpha+\beta +T_y n-m}{\alpha-\beta+m-n}\right)_x,
\end{array}
\end{equation*}
which is equivalent to
\begin{equation}\label{Case3disp}
\begin{array}{c}
\triangle_t (\alpha+m)=\left(\ln \frac{T_t \alpha+\beta +T_t m-n}{\alpha-\beta+m-n}\right)_x,\quad
\triangle_y (\beta+n)=\left(\ln \frac{T_t \alpha+\beta +T_y n-m}{\alpha-\beta+m-n}\right)_x,\\
\triangle_t \alpha=\triangle_y \beta=\frac{1}{2}\left(\ln \frac{(T_t \alpha+\beta +T_t m-n)(T_t \alpha+\beta +T_y n-m)}{(\alpha-\beta+m-n)(\alpha-\beta+T_y n-T_t m)}\right)_x.
\end{array}
\end{equation}
Dispersionless Lax pair (\ref{lax3}) follows from  dispersive Lax equations  (\ref{Lax3}) via the standard substitution  $\psi=e^{S/\epsilon}$,
followed by cancelling the common factor  $e^{S/\epsilon}$ and taking the limit $\epsilon\to 0$. Here the first two Lax equations (\ref{Lax3}) give  dispersionless equations (\ref{lax3}), while the third equation (\ref{Lax3}) gives no additional constraints for $S$; note that the $\psi$-term in the expanded form of third equation (\ref{Lax3}) can be set equal to zero.
The corresponding dispersive deformation of Lagrangian (\ref{ZL3}) is
\begin{equation}\label{ZL3disp}
\int \left((\alpha+\frac{\epsilon}{2}(\triangle_t \alpha+ \triangle_y n))\, \partial_x^{-1}\triangle_t m
-(\beta+\frac{\epsilon}{2}\triangle_t \alpha)\, \partial_x^{-1}\triangle_y n-\tilde{l}\right)\, dx\delta y \delta t
\end{equation}
where
$$
\tilde{l}=\varphi(T_t \alpha+\beta +T_t m-n)+\varphi(\alpha-\beta+T_y n-T_t m)-\varphi(\alpha-\beta+m-n)-\varphi(T_t \alpha+\beta +T_y n-m).
$$
Here integration over the discrete variables $y$ and $t$, indicated by $\delta y$ and $\delta t$, is understood as  summation over all $T_y$ and $T_t$-translates of the Lagrangian density (\ref{ZL3disp}).
The last two equations of system (\ref{Case3disp}) suggest the introduction of a potential $u$ such that
\begin{equation}\label{potu3}
\alpha=\triangle_y u_{x}, \quad \beta=\triangle_t u_{x}, \quad
\ln \frac{(T_t \alpha+\beta +T_t m-n)(T_t \alpha+\beta +T_y n-m)}{(\alpha-\beta+m-n)(\alpha-\beta+T_y n-T_t m)} =2\triangle_y \triangle_t u.
\end{equation}
 In what follows, we will use the quantity $q$ defined by
 \begin{equation}\label{q3}
q^2-\chi q+\zeta=0,
\end{equation}
where
$$
\chi=(\triangle_y u_x+\triangle_t u_x)(1-e^{-2\triangle_y\triangle_t u})
-\epsilon \triangle_y\triangle_t u_x e^{-2\triangle_y\triangle_t u},\quad \zeta=\triangle_y u_x\triangle_t u_x(1-e^{-2\triangle_y\triangle_t u} ).
$$


\begin{proposition} \label{pc3} System (\ref{Case3disp}) can be written as a single differential-difference equation for the potential  $u$,
\begin{equation}\label{C3D}
\begin{array}{c}
\triangle_t\left(\ln(\frac{q}{\triangle_t u_x}-1)+\triangle_y\triangle_t u\right)_x
+\triangle_y \left(\ln(1-\frac{q}{\triangle_y u_x}) +\triangle_y\triangle_t  u  \right)_x\\
\ \\
+\triangle_y\triangle_t\left(
{\triangle_t u_x}+{\triangle_y u_x}-\frac{2q}{ 1-e^{-2\triangle_y\triangle_t  u}}
\right)
 -
\triangle_y\triangle_t\left(\epsilon
\ln \frac{ \triangle_y u_x \triangle_t u_x (1-e^{-2\triangle_y\triangle_t  u})}{q}
\right)_x=0.
\end{array}
\end{equation}
Equation (\ref{C3D}) is represented in Euler-Lagrange form corresponding to a  Lagrangian $\int F\, dx\delta y\delta t$, with the Lagrangian density
\begin{equation}\label{f3disp}
\begin{array}{c}
F=
\triangle_t u_x\ln(\frac{q}{\triangle_t u_x}-1)+\triangle_y u_x  \ln(1-\frac{q}{\triangle_y u_x})+({\triangle_t u_x}+{\triangle_y u_x})\triangle_y\triangle_t u
\\
+\epsilon \triangle_y\triangle_t u_x \ln(q+\epsilon\triangle_y\triangle_t u_x)-\epsilon \triangle_y\triangle_t u_x.
\end{array}
\end{equation}
In the dispersionless limit $\epsilon \to 0$,
we have $q=\frac{2u_{xy}u_{xt}}{u_{xy}+u_{xt}+L}$, where $L ={\sqrt{u_{xt}^{2}+u_{xy}^{2}-2u_{xt}u_{xy}\coth u_{yt}}}$, and the dispersive Lagrangian density $F$ reduces to the dispersionless Lagrangian density $f$
given by (\ref{f3hyp}).

\end{proposition}

\noindent {\it Proof:}
Parametrising the last relation of \eqref{potu3} in the form
\begin{equation}\label{Case3disp1}
\begin{array}{c}
     \frac{\beta-\alpha+n-m}{T_t \alpha+\beta +T_t m-n}=\xi,\quad
 \frac{\alpha-\beta+m-n}{T_t \alpha+\beta +T_y n-m}=\eta,\quad
 \frac{\beta-\alpha+n-m}{\alpha-\beta+T_y n-T_t m} =\xi \eta e^{2\triangle_y \triangle_t u},
\end{array}
\end{equation}
and substituting into the first two  equations of system (\ref{Case3disp}), one obtains
\begin{equation}\label{Case3disp20}
\begin{array}{c}
  \triangle_t m=-(\ln(-\xi)+\triangle_y\triangle_t  u)_x, \quad  \triangle_y n=-(\ln\eta+\triangle_y\triangle_t  u)_x.
\end{array}
\end{equation}
Then \eqref{Case3disp1} yields
\begin{equation}\label{Case3disp200}
\begin{array}{c}
m-n+{\triangle_t u_x}+{\triangle_y u_x}-\frac{2}{1+\xi} {\triangle_t u_x}-
(\epsilon\ln(1+\xi))_x=0,\\
\ \\
{\triangle_t u_x}+{\triangle_y u_x} -\frac{1}{1+\xi} {\triangle_t u_x}-\frac{1}{1+\eta} {\triangle_y u_x}
-\frac{e^{-2\triangle_y\triangle_t u}}{1-e^{-2\triangle_y\triangle_t u} }({\epsilon\triangle_y\triangle_t u_x })=0,\\
\ \\
\left(1+{\eta}\right)\left(1+{\xi}\right)=1-e^{-2\triangle_y\triangle_t u} .
\end{array}
\end{equation}
Applying to the first equation of \eqref{Case3disp200}  the operator  $\triangle_y \triangle_t $ and using  \eqref{Case3disp20},   we obtain
\begin{equation}\label{Case3pde}
\begin{array}{c}
\triangle_t(\ln\eta+\triangle_y\triangle_t u)_x
+\triangle_y (-\ln(-\xi)-\triangle_y\triangle_t  u)_x\\
\ \\
+\triangle_y\triangle_t\left(
{\triangle_t u_x}+{\triangle_y u_x}-\frac{2}{1+\xi} {\triangle_t u_x}-
(\epsilon\ln(1+\xi))_x
\right)=0.
\end{array}
\end{equation}
To solve for $\xi, \ \eta$, we
introduce new variable $q$ such that $\eta=\frac{q}{\triangle_t u_x}-1$.
Then the last two equations of \eqref{Case3disp200} lead to  relation \eqref{q3}.
The last equation of \eqref{Case3disp200} gives
\begin{equation*}
 \xi=\frac{\triangle_t u_x (1-e^{-2\triangle_y\triangle_t  u})}{q}-1,
\end{equation*}
which can also be written as
\begin{equation*}
 \xi=-\frac{\triangle_y u_x}{\triangle_y u_x-q}e^{-2\triangle_y\triangle_t u +\epsilon \triangle_t\ln \triangle_y u_x},
\end{equation*}
by using  relation \eqref{q3}. Inserting the expressions of $\xi, \eta$ into \eqref{Case3pde} results in equation \eqref{C3D}.
To reconstruct the corresponding Lagrangian, note that for a Lagrangian density of the form   $F=F(\triangle_yu_x,\ \triangle_tu_x,\ \triangle_y\triangle_tu,\ \triangle_y\triangle_tu_{x})$, the  Euler-Lagrange equation is
$$
\begin{array}{c}
\triangle_y\left(\frac{\partial F}{\partial (\triangle_yu_{x})}\right)_{x}+\triangle_t\left(\frac{\partial F}{\partial (\triangle_tu_{x})}\right)_{x}+\triangle_y\triangle_t\left(\frac{\partial F}{\partial (\triangle_y\triangle_t u)}\right)\\
-\triangle_y\triangle_t\left(\frac{\partial F}{\partial (\triangle_y\triangle_tu_{x})}-\epsilon\frac{\partial F}{\partial (\triangle_yu_{x})}-\epsilon\frac{\partial F}{\partial (\triangle_tu_{x})}\right)_x=0.
\end{array}
$$
Comparison with (\ref{C3D}) gives the expressions for all first-order derivatives of $F$:
$$
 \frac{\partial F}{\partial (\triangle_y u_{x})}=\ln (1-\frac{q}{\triangle_y u_x}) +\triangle_y\triangle_t u, \quad
  \frac{\partial F}{\partial (\triangle_tu_{x})}= \ln (\frac{q}{\triangle_t u_x}-1) +\triangle_y\triangle_t u ,
  $$
  $$
   \frac{\partial F}{\partial (\triangle_y\triangle_t u)}= {\triangle_t u_x}+{\triangle_y u_x}-\frac{2q}{ 1-e^{-2\triangle_y\triangle_t u}},
   $$
   $$
    \frac{\partial F}{\partial (\triangle_y\triangle_t u_x)}=\epsilon
\ln \frac{ (1-e^{-2\triangle_y\triangle_t u})(\triangle_y u_x-q)(q-\triangle_t u_x)}{q}+2\epsilon \triangle_y \triangle_t u.
$$
By using \eqref{q3},  the expression of $\frac{\partial F}{\partial (\triangle_y\triangle_t u_x)} $ can be simplified as
  $$
    \frac{\partial F}{\partial (\triangle_y\triangle_t u_x)}=\epsilon
\ln (q+\epsilon \triangle_y\triangle_t u_x ).
$$
One can show that this system for $F$ is consistent and, on integration, leads to the Lagrangian density \eqref{f3disp}.
In the dispersionless limit $\epsilon \to 0$, we have $q=\frac{2u_{xy}u_{xt}}{u_{xy}+u_{xt}+L}$ where $L =\sqrt{{u_{xt}^{2}+u_{xy}^{2}-2u_{xt}u_{xy}\coth u_{yt}}}$,
 and the dispersive Lagrangian density $F$ reduces to
\begin{equation*}
\begin{array}{c}
f=
u_{xt}\ln(\frac{q}{u_{xt}}-1)+u_{xy}  \ln(1-\frac{q}{u_{xy}} )+(u_{xt} +u_{xy} )u_{yt}
 \\
 \\
=\frac{u_{xt}+u_{xy}}{2}\ln \left( {(\frac{q}{u_{xt}}-1)}{  (1-\frac{q}{u_{xy}} )e^{2u_{yt}}} \right)
+\frac{u_{xt}-u_{xy}}{2}\ln \left(\frac{\frac{q}{u_{xt}}-1}{  1-\frac{q}{u_{xy}} }\right)
 \\
 \\
=\frac{u_{xt}+u_{xy}}{2}\ln \left(\frac{(u_{xy}+u_{xt})q}{u_{xy}u_{xt}}-1\right)
+\frac{u_{xt}-u_{xy}}{2}\ln \left(\frac{\frac{q}{u_{xt}}-1}{  1-\frac{q}{u_{xy}}}\right)
\\
\\=\frac{1}{2}({u_{xt}+u_{xy}})\ln \left(\frac{u_{xy}+u_{xt}-L}{u_{xy}+u_{xt}+L}\right)
-\frac{1}{2}({u_{xy}-u_{xt}})\ln\left( \frac{{u_{xy}}-u_{xt}-L}{  u_{xy}-u_{xt}+L }\right),
\end{array}
\end{equation*}
which is
exactly (\ref{f3hyp}).
Modulo total derivatives and rescalings, the Lagrangian density \eqref{f3disp} is equivalent to the one constructed in \cite{XFP} for the semi-discrete Darboux system with two discrete variables.
$\square$


\subsection{Case 4 }
\label{sec:Case4}

Here we use formula \eqref{hyphex} for $f$. The corresponding dispersionless Lax pair has the general form
 \begin{equation}\label{lax4}
S_y=\ln{\frac{e^{S_x+\alpha +m}-1}{e^{S_x+m}-e^{\alpha}}},
\quad
 S_t=\ln{\frac{e^{S_x+\beta +n}-1}{e^{S_x+n}-e^{\beta}}}.
\end{equation}
The compatibility condition $S_{yt}=S_{ty}$ gives the  dispersionless equations,
\begin{equation}\label{Case4}
\begin{array}{c}
m_t=\frac{1}{2}\left(\ln \frac{(e^{\alpha+\beta +m}-e^{n})(e^{\alpha+n}-e^{\beta+m}) }{(e^{\alpha+\beta +n}-e^{m})(e^{\alpha+m}-e^{\beta+n})}\right)_x,  \quad
n_y=\frac{1}{2}\left(\ln \frac{(e^{\alpha+\beta +n}-e^{m})(e^{\alpha+n}-e^{\beta+m}) }{(e^{\alpha+\beta +m}-e^{n})(e^{\alpha+m}-e^{\beta+n})}\right)_x, \\
\ \\
\alpha_t=\beta_y=\frac{1}{2}\left(\ln \frac{(e^{\alpha+\beta +m}-e^{n})(e^{\alpha+\beta +n}-e^{m})}{(e^{\alpha+m}-e^{\beta+n})(e^{\alpha+n}-e^{\beta+m})}\right)_x,
\end{array}
\end{equation}
equivalently,
\begin{equation}\label{Case44}
\begin{array}{c}
(\alpha+m)_t=\left(\ln \frac{e^{\alpha+\beta+m}-e^{n} }{e^{\alpha+m}-e^{\beta+n}}\right)_x,  \quad
(\beta+n)_y=\left(\ln \frac{e^{\alpha+\beta+n}-e^{m}}{e^{\alpha+m}-e^{\beta+n}}\right)_x, \\
\ \\
\alpha_t=\beta_y=\frac{1}{2}\left(\ln \frac{(e^{\alpha+\beta+m}-e^{n})(e^{\alpha+\beta+n}-e^{m})}{(e^{\alpha+m}-e^{\beta+n})(e^{\alpha+n}-e^{\beta+m})}\right)_x.
\end{array}
\end{equation}
System (\ref{Case4}) can also be written in the form
\begin{equation}\label{Case42}
\begin{array}{c}
m_t=\frac{1}{2}\left(\ln \frac{\sinh\frac{\alpha+\beta +m-n}{2}\sinh\frac{\beta-a+m-n}{2} }{\sinh\frac{\alpha+\beta +n-m}{2}\sinh\frac{\alpha-\beta+m-n}{2}}\right)_x,  \quad
n_y=\frac{1}{2}\left(\ln \frac{\sinh\frac{\alpha+\beta +n-m}{2}\sinh\frac{\beta-a+m-n}{2} }{\sinh\frac{\alpha+\beta +m-n}{2}\sinh\frac{\alpha-\beta+m-n}{2}}\right)_x, \\
\ \\
\alpha_t=\beta_y=\frac{1}{2}\left(\ln \frac{\sinh\frac{\alpha+\beta +m-n}{2}\sinh\frac{\alpha+\beta +n-m}{2}}{\sinh\frac{\alpha-\beta+m-n}{2}\sinh\frac{\beta-a+m-n}{2}}\right)_x,
\end{array}
\end{equation}
which comes from the Lagrangian
\begin{equation}\label{ZL4}
\int \left(\alpha\, \partial_x^{-1}m_t-{\beta}\, \partial_x^{-1}n_y+l\right)\, dxdydt
\end{equation}
where
$$
\begin{array}{c}
l=\Gamma\left(\frac{\alpha+\beta +m-n}{2}\right)-\Gamma\left(\frac{\alpha+\beta +n-m}{2}\right)-\Gamma\left(\frac{\beta-a+m-n}{2}\right)-\Gamma\left(\frac{\alpha-\beta+m-n}{2}\right);
\end{array}
$$
here $\Gamma(s)=-\int_0^s \ln \sinh\xi \, d \xi$ is yet another hyperbolic version of the Lobachevsky function.
Under the substitution $\alpha=v_3 , \, \beta=v_2, \, m=p, \,n=p-f_{v_1}$,
dispersionless Lax pair (\ref{lax4}) goes to (\ref{c4hyp}), the last two equations (\ref{Case4}) will be satisfied identically, while the first two equations (\ref{Case4})  lead to equations (\ref{pdef}) that are equivalent to the Euler-Lagrange equation for the corresponding Lagrangian density $f$.

To construct dispersive deformation of dispersionless Lax pair (\ref{lax4}), we first rewrite it in equivalent form,
\begin{equation*}
\begin{array}{ccc}
\\
e^{ S_y+S_x}=e^{\alpha+S_x}+e^{\alpha-m+S_y}-e^{-m},\\
 e^{S_t+S_x}=e^{\beta+S_x}+e^{\beta-n+S_t}-e^{-n},\\
e^{S_y+S_t}=\frac{e^{\alpha+\beta -m}-e^{-n}}{e^{\alpha-m}-e^{\beta-n}}e^{S_y}-\frac{e^{\alpha+\beta -n}-e^{-m}}{e^{\alpha-m}-e^{\beta-n}}e^{S_t}
+\frac{e^{\alpha-n}-e^{\beta-m}}{e^{\alpha-m}-e^{\beta-n}},
\end {array}
\end{equation*}
where the third equation is an algebraic consequence of the first two. Its dispersive deformation  is
\begin{equation}\label{Lax4}
\begin{array}{ccc}
T_yT_x\psi=e^{\alpha}T_x\psi+e^{\alpha-m}T_y\psi -e^{-m}\psi, \\
\ \\
T_tT_x\psi=e^{\beta}T_x\psi+e^{\beta-n}T_t\psi -e^{-n}\psi, \\
\ \\
T_yT_t\psi=\frac{e^{\alpha+T_y \beta-m}-e^{-T_y n}}{e^{T_t (\alpha-m)}-e^{T_y(\beta-n)}}T_y \psi
-\frac{e^{T_t \alpha+\beta -n}-e^{-T_t m}}{e^{T_t (\alpha-m)}-e^{T_y(\beta-n)}}T_t \psi
+\frac{e^{T_t \alpha-n}-e^{T_y \beta-m}}{e^{T_t (\alpha-m)}-e^{T_y(\beta-n)}}\psi.
\end{array}
\end{equation}
The compatibility conditions of dispersive Lax equations (\ref{Lax4}) can be written in several equivalent forms.
One of them is
\begin{equation}\label{Case4d0}
\begin{array}{c}
\triangle_t \alpha=\triangle_y\beta, \\ \\
\triangle_t (\alpha-m)=\triangle_x \left(\ln  \frac{e^{T_y \beta+\alpha -m}-e^{-T_yn} }{e^{T_t(\alpha-m)}-e^{T_y(\beta-n)}}\right),  \\ \\
\triangle_y (\beta-n)= \triangle_x \left( \ln  \frac{e^{T_t \alpha+\beta-n}-e^{-T_tm} }{e^{T_t(\alpha-m)}-e^{T_y(\beta-n)}}\right), \\ \\
\triangle_t m+\triangle_y n=-\triangle_x \left( \ln  \frac{e^{T_t \alpha-n}-e^{T_y \beta-m} }{e^{T_t(\alpha-m)}-e^{T_y(\beta-n)}}\right).
\end{array}
\end{equation}
An equivalent form, from which dispersionless limit  (\ref{Case44}) can be most easily seen, is:
\begin{equation}\label{Case4d}
\begin{array}{c}
T_x\triangle_t (\alpha+m)=\triangle_x \left(\ln  \frac{e^{T_t \alpha+\beta +T_t m}-e^{n} }{e^{\alpha+m}-e^{\beta+n}}\right),  \\ \\
T_x\triangle_y (\beta+n)= \triangle_x \left(\ln \frac{e^{T_y \beta+\alpha +T_y n}-e^{m} }{e^{\beta+n}-e^{\alpha+m}}\right), \\
\ \\
T_x \triangle_t \alpha=T_x \triangle_y \beta=\frac{1}{2}\triangle_x\left(\ln \frac{(e^{T_t \alpha+\beta +T_t m}-e^{n})(e^{T_y \beta+\alpha +T_y n}-e^{m})}{(e^{\beta+n}-e^{\alpha+m})(e^{\beta+T_t m}-e^{\alpha+T_y n})}\right),
\end{array}
\end{equation}
which is also equivalent to
\begin{equation}\label{Case4disp}
\begin{array}{c}
\triangle_t (\alpha+m)+\frac{1}{2}\epsilon\triangle_t\triangle_x (\alpha+m)
=\triangle_x\left(
\ln \frac{\sinh\frac{T_t \alpha+\beta +T_t m-n}{2}}{\sinh\frac{\alpha-\beta+m-n}{2}}
\right), \\
\\
\triangle_y (\beta+n)+\frac{1}{2}\epsilon\triangle_y\triangle_x (\beta+n)=\triangle_x\left(\ln \frac{\sinh\frac{ T_y \beta+\alpha +T_yn-m}{2}}{\sinh\frac{\alpha-\beta+m-n}{2}}\right),\\
\\
\
\triangle_t \alpha=\triangle_y \beta
=-\frac{1}{2}\epsilon\triangle_t\triangle_x \alpha+
\frac{1}{2}\triangle_x\left(
\ln \frac{\sinh\frac{T_t \alpha+\beta +T_t m-n}{2}\sinh\frac{ T_y \beta+\alpha +T_yn-m}{2}
}{\sinh\frac{\alpha-\beta+m-n}{2}\sinh\frac{\beta-\alpha+T_t m-T_y n}{2}}
\right).
\end{array}
\end{equation}
Dispersionless Lax pair (\ref{lax4}) follows from  dispersive Lax equations  (\ref{Lax4}) via the standard substitution  $\psi=e^{S/\epsilon}$,
followed by cancelling the common factor  $e^{S/\epsilon}$ and taking the limit $\epsilon\to 0$. Here the first two Lax equations (\ref{Lax4}) give  dispersionless equations (\ref{lax4}), while the third equation (\ref{Lax4}) gives no additional constraints for $S$.
The corresponding dispersive deformation of Lagrangian (\ref{ZL4}) is
\begin{equation}\label{ZL4disp}
\int \left(
(\alpha+\frac{\epsilon}{2}\triangle_t \alpha- \frac{1}{2}\triangle_y\triangle_x^{-1} n)C
-(\beta+\frac{\epsilon}{2}\triangle_y \beta- \frac{1}{2}\triangle_t\triangle_x^{-1} m)B
+\tilde{l}
\right)\, \delta x\delta y \delta t.
\end{equation}
Here  integration over the discrete variables $x, y, t$, denoted $ \int \delta x \delta y \delta t$, is understood as summation over all $x, y, t$-translates of the  Lagrangian density (\ref{ZL4disp}); the quantities
$C, \ B, \ \tilde{l}$ are defined as
$$
C=\triangle_t\triangle_x^{-1} m+\frac{\epsilon \triangle_t m}{2},\quad  B=\triangle_y\triangle_x^{-1} n+\frac{\epsilon \triangle_y n}{2},
$$
$$
\begin{array}{c}
\tilde{l}=\Gamma \left(\frac{T_t \alpha+\beta +T_t m-n}{2}\right)-\Gamma \left(\frac{T_t \alpha+\beta +T_y n-m}{2}\right)-\Gamma \left(\frac{\beta-\alpha+T_t m-T_y n}{2}\right)
-\Gamma \left(\frac{\alpha-\beta+m-n}{2}\right).
\end{array}
$$
The last two equations of system (\ref{Case4d}) suggest the introduction of a potential $u$ such that
\begin{equation}\label{potu4}
\alpha=\triangle_x\triangle_y u, \quad \beta=\triangle_x\triangle_t u, \quad
\frac{1}{2}\ln\frac{(e^{T_t \alpha+\beta +T_t m}-e^{n})(e^{T_y \beta+\alpha +T_y n}-e^{m})}{(e^{\beta+n}-e^{\alpha+m})(e^{\beta+T_t m}-e^{\alpha+T_y n})} =T_x \gamma,
\end{equation}
where $\gamma=\triangle_y \triangle_t u$.
 In what follows, we will use the quantities $ \widehat{\alpha}, \ \widehat{\beta}, \ \widehat{\gamma}, \ \widehat{u}$ defined by
\begin{equation}\label{c4a9}
 \widehat{\alpha}=\triangle_x\triangle_y \widehat{u}, \quad \widehat{\beta}=\triangle_x\triangle_t \widehat{u}, \quad \widehat{\gamma}=\triangle_y\triangle_t \widehat{u},\quad \widehat{u}=-(1+T_x^{-1})u,
 \end{equation}
 as well as the quantity $q$ such that
 \begin{equation}\label{q4}
q^2-\chi q+\zeta=0,
\end{equation}
where
$$
\chi=e^{\widehat{\alpha}}+e^{\widehat{\beta}}+e^{\widehat{\gamma}}-e^{\widehat{\alpha}+\widehat{\beta}+\widehat{\gamma}+\epsilon \triangle_x \widehat{\gamma}}-2,
\quad \zeta=(1-e^{\widehat{\alpha}})(1-e^{\widehat{\beta}})(1-e^{\widehat{\gamma}}).
$$

\begin{proposition} \label{pc4} System (\ref{Case4disp}) can be written as a single difference equation for $\widehat{u}$,
 \begin{equation}\label{c4PDE}
\begin{array}{c}
 \triangle_x\triangle_y \left(\ln\left(1+\frac{q}{1-e^{\widehat{\alpha}}}\right)-\frac{1}{2}\widehat{\beta}-\frac{1}{2}\widehat{\gamma}\right)
+\triangle_x\triangle_t \left(\ln\left(1+\frac{q}{1-e^{\widehat{\beta}}}\right)-\frac{1}{2}\widehat{\alpha}-\frac{1}{2}\widehat{\gamma}\right)
\\
+\triangle_y\triangle_t \left(\ln\left(1+\frac{q}{1-e^{\widehat{\gamma}}}\right)-\frac{1}{2}\widehat{\alpha}-\frac{1}{2}\widehat{\beta}\right)
-\epsilon \, \triangle_x\triangle_y\triangle_t \ln\frac{(1-e^{\widehat{\alpha}})(1-e^{\widehat{\beta}})(1-e^{\widehat{\gamma}})}{q}=0.
\end{array}
\end{equation}
Equation (\ref{c4PDE}) is represented in Euler-Lagrange form corresponding to a Lagrangian $ \int F\, \delta x\delta y\delta t$, with the Lagrangian density
\begin{equation}\label{c4F}
\begin{array}{c}
F=
 \operatorname{Li}_2\left(e^{\widehat{\alpha}}\right)
+ \operatorname{Li}_2\left(e^{\widehat{\beta}}\right)
+ \operatorname{Li}_2\left(e^{\widehat{\gamma}}\right)
+ \operatorname{Li}_2\left(\frac{1}{1+q}\right)
\ \\
-\operatorname{Li}_2\left(\frac{e^{\widehat{\alpha}}}{1+q}\right)
-\operatorname{Li}_2\left(\frac{e^{\widehat{\beta}}}{1+q}\right)
-\operatorname{Li}_2\left(\frac{e^{\widehat{\gamma}}}{1+q}\right)
-\operatorname{Li}_2\left((1+q) e^{\epsilon\triangle_x \widehat{\gamma}} \right)
\ \\
+(\widehat{\alpha}+\widehat{\beta}+\widehat{\gamma}-\ln(1+q))\ln(1+q)
-\frac{1}{2}(\widehat{\alpha}\widehat{\beta}+\widehat{\alpha}\widehat{\gamma} +\widehat{\beta}\widehat{\gamma}),
\end{array}
\end{equation}
where
the dilogarithm function is defined as $\operatorname{Li}_2(x)=-\int_0^x \frac{\ln (1-t)}{t} dt$. In the dispersionless limit $\epsilon \to 0$,  modulo a constant, the dispersive Lagrangian density $-F/2$ reduces to the dispersionless Lagrangian density $f$ given by 
\eqref{fHH}.
\end{proposition}

 \noindent {\it Proof:}
Parametrising the last relation of \eqref{potu4} in the form
\begin{equation}\label{Case4disp1}
\begin{array}{c}
   \frac{e^{\beta+n}-e^{\alpha+m}}{e^{T_t\alpha+\beta +T_t m}-e^{n} }=e^\beta T_x\xi,\quad
  \frac{e^{\alpha+m}-e^{\beta+n}}{e^{T_y\beta+\alpha +T_y n}-e^{m} }=e^\alpha T_x\eta,\quad
 \frac{e^{\alpha+m}-e^{\beta+n}}{e^{\beta+T_t m}-e^{\alpha+T_y n}} =e^{\alpha+\beta }T_x(\xi \eta e^{2\gamma}),
\end{array}
\end{equation}
and substituting into the first two equations of  system (\ref{Case4d}), one obtains
\begin{equation}\label{Case4disp20}
\begin{array}{c}
\triangle_t m=-\triangle_t \alpha-\triangle_x \left( T_x^{-1} \beta+ \ln(-\xi)\right), \quad  \triangle_y n=-\triangle_y \beta-\triangle_x \left( T_x^{-1} \alpha+\ln (-\eta)\right).
\end{array}
\end{equation}
Then  \eqref{Case4disp1} yield
\begin{equation}\label{Case4disp200}
\begin{array}{c}
(m+\alpha )-(n+\beta )-\widehat{\beta}+\ln\left( 1+\frac{e^{\widehat{\beta}}-1}{1+\xi}\right) -\triangle_x(\epsilon\ln(1+\xi))=0,\\
1-e^{T_x\widehat{\gamma}}=(1+e^{-\widehat{\beta}} \xi)(1+e^{-\widehat{\alpha}} \eta),
\ \\
T_x(\left(1+{\eta}\right)\left(1+{\xi}\right))=T_x(1-e^{\widehat{\gamma}}),
\end{array}
\end{equation}
where $\widehat{\alpha}, \ \widehat{\beta}, \ \widehat{\gamma}$ are as in \eqref{c4a9}.
Applying to the first equation of \eqref{Case4disp200}  the operator  $\triangle_y \triangle_t $ and using  \eqref{Case4disp20},   we obtain
\begin{equation}\label{Case4pde}
\begin{array}{c}
 \triangle_x\triangle_t\left(\ln (-\eta)\right)
+\triangle_x\triangle_y  \left( - \ln(-\xi)\right)\\
+\triangle_y\triangle_t\left(-\widehat{\beta}+\ln\left( 1+\frac{e^{\widehat{\beta}}-1}{1+\xi}\right)\right)
-\triangle_x\triangle_y\triangle_t\left(\epsilon\ln(1+\xi)
\right)=0.
\end{array}
\end{equation}
Introducing the new variable $q$ such that $\eta={q}/{(e^{\widehat{\beta}}-1)}-1$, the last equation of \eqref{Case4disp200} gives
$
 \xi=-{(1-e^{\widehat{\gamma}})(1-e^{\widehat{\beta}})}/{q}-1,
$
then the second equation of \eqref{Case4disp200} leads to the relation \eqref{q4}.
It is noted that $\xi$ can also be written as
\begin{equation*}
 \quad
 \xi=-\frac{1-e^{\widehat{\alpha}}}{1-e^{\widehat{\alpha}}+q}
 e^{\widehat{\beta}+\widehat{\gamma}+\epsilon \triangle_t\ln (e^{\widehat{\alpha}}-1) }.
\end{equation*}
Inserting the expressions of $\xi, \eta$ into \eqref{Case4pde} results in
\begin{equation*}
\begin{array}{c}
 \triangle_x\triangle_t\ln\left(1+\frac{q}{1-e^{\widehat{\beta}}}\right)
+\triangle_x\triangle_y \left(-\widehat{\beta}-\widehat{\gamma}+\ln\left(1+\frac{q}{1-e^{\widehat{\alpha}}}\right)\right)
\\
+\triangle_y\triangle_t\left(-\widehat{\beta}+\ln\left(1+\frac{q}{1-e^{\widehat{\gamma}}}\right)\right)
-\triangle_x\triangle_y\triangle_t\left(\epsilon\ln\frac{(1-e^{\widehat{\alpha}})(1-e^{\widehat{\beta}})(1-e^{\widehat{\gamma}})}{q}
\right)=0,
\end{array}
\end{equation*}
which is equivalent to \eqref{c4PDE}.
To reconstruct the corresponding Lagrangian, note that for a Lagrangian density of the form  $F=F(\triangle_x\triangle_y\widehat{u}, \ \triangle_x\triangle_t\widehat{u},\  \triangle_y\triangle_t\widehat{u}, \ \triangle_x\triangle_y\triangle_t\widehat{u})$, the  Euler-Lagrange equation is
$$
\begin{array}{c}
\triangle_x\triangle_y\left(\frac{\partial F}{\partial (\triangle_x\triangle_y\widehat{u})}\right)+\triangle_x\triangle_t\left(\frac{\partial F}{\partial (\triangle_x \triangle_t\widehat{u})}\right)+\triangle_y\triangle_t\left(\frac{\partial F}{\partial (\triangle_y\triangle_t \widehat{u})}\right)\\
-\triangle_x\triangle_y\triangle_t\left(\frac{\partial F}{\partial (\triangle_x\triangle_y\triangle_t\widehat{u})}-\epsilon\frac{\partial F}{\partial (\triangle_x\triangle_y\widehat{u})}-\epsilon\frac{\partial F}{\partial (\triangle_x\triangle_t\widehat{u})}-\epsilon\frac{\partial F}{\partial (\triangle_y\triangle_t\widehat{u})}\right)=0.
\end{array}
$$
Comparison with (\ref{c4PDE}) gives the expressions for all first-order derivatives of $F$:
$$
 \frac{\partial F}{\partial \widehat{\alpha}}=\ln\left(1+\frac{q}{1-e^{\widehat{\alpha}}}\right)-\frac{1}{2}\widehat{\beta}-\frac{1}{2}\widehat{\gamma}, \quad
  \frac{\partial F}{\partial \widehat{\beta}}= \ln\left(1+\frac{q}{1-e^{\widehat{\beta}}}\right)-\frac{1}{2}\widehat{\alpha}-\frac{1}{2}\widehat{\gamma},
  $$
  $$
   \frac{\partial F}{\partial \widehat{\gamma}}= \ln\left(1+\frac{q}{1-e^{\widehat{\gamma}}}\right)-\frac{1}{2}\widehat{\alpha}-\frac{1}{2}\widehat{\beta},
   $$
   $$
    \frac{\partial F}{\partial (\triangle_x \widehat{\gamma} )}=\epsilon
\ln \frac{(1-e^{\widehat{\alpha}}+q)(1-e^{\widehat{\beta}}+q)(1-e^{\widehat{\gamma}}+q)}{q}-\epsilon(\widehat{\alpha}+\widehat{\beta}+\widehat{\gamma}).
$$
Using \eqref{q4},  the expression for $\frac{\partial F}{\partial(\triangle_x \widehat{\gamma} )} $ can be simplified as
  $$
    \frac{\partial F}{\partial (\triangle_x \widehat{\gamma} )}=\epsilon
\ln (1-(1+q)e^{\epsilon\triangle_x \widehat{\gamma}} ).
$$
One can show that this system for $F$ is consistent and, on integration, leads to the Lagrangian density \eqref{c4F}.
In the dispersionless limit $\epsilon \to 0$,  we have
\begin{equation*}
 \widehat{\alpha}= \widehat{u}_{xy}, \quad \widehat{\beta}=\widehat{u}_{xt}, \quad \widehat{\gamma}= \widehat{u}_{yt},\quad \widehat{u}=-2u,
 \end{equation*}
 \begin{equation*}
q=\frac{\chi-L}{2},\quad L^2={\chi^2-4\zeta},
\end{equation*}
where
$$
\chi=e^{\widehat{\alpha}}+e^{\widehat{\beta}}+e^{\widehat{\gamma}}-e^{\widehat{\alpha}+\widehat{\beta}+\widehat{\gamma}}-2,
\quad \zeta=(1-e^{\widehat{\alpha}})(1-e^{\widehat{\beta}})(1-e^{\widehat{\gamma}}),
$$
 and the dispersive Lagrangian density $-F/2$ reduces to
\begin{equation}\label{f4ln}
\begin{array}{c}
 f=
-\frac{1}{2}  \operatorname{Li}_2\left(e^{-2u_{xy}}\right)
-\frac{1}{2}\operatorname{Li}_2\left(e^{-2u_{xt}}\right)
-\frac{1}{2} \operatorname{Li}_2\left(e^{-2u_{yt}}\right)
-\frac{1}{2} \operatorname{Li}_2\left(\frac{1}{1+q}\right)
\ \\
+\frac{1}{2}\operatorname{Li}_2\left(\frac{e^{-2u_{xy}}}{1+q}\right)
+\frac{1}{2}\operatorname{Li}_2\left(\frac{e^{-2u_{xt}}}{1+q}\right)
+\frac{1}{2}\operatorname{Li}_2\left(\frac{e^{-2u_{yt}}}{1+q}\right)
+\frac{1}{2}\operatorname{Li}_2\left(1+q \right)
\ \\
+\frac{1}{2}\left(\ln(1+q)+2u_{xy}+2u_{xt}+2u_{yt}\right)\ln(1+q)
+u_{xy}u_{xt}+u_{xy}u_{yt} +u_{xt}u_{yt},
\end{array}
\end{equation}
with  first-order derivatives
$$
 \frac{\partial f}{\partial u_{xy}}=\ln\left(1+\frac{q}{1-e^{-2u_{xy}}}\right)+u_{xt}+u_{yt}, \quad
  \frac{\partial f}{\partial u_{xt}}= \ln\left(1+\frac{q}{1-e^{-2u_{xt}}}\right)+u_{xy}+u_{yt},
  $$
  $$
   \frac{\partial f}{\partial u_{yt}}= \ln\left(1+\frac{q}{1-e^{-2u_{yt}}}\right)+u_{xy}+u_{xt}.
   $$
Assuming $u_{xy}<0, u_{xt}<0, u_{yt}<0, \  L=\sqrt{\chi^2-4\zeta}$, the  first-order derivatives can be written as
\begin{equation*}
\begin{array}{c}
 \frac{\partial f}{\partial u_{xy}}=\arccosh\left(\frac{\coth{(u_{xt})}\coth{(u_{yt})}-\coth{(u_{xy})}}{{\rm csch}{(u_{xt})}{\rm csch}{ (u_{yt})}}\right),\\
   \frac{\partial f}{\partial u_{xt}}=\arccosh\left(\frac{\coth{(u_{xy})}\coth{(u_{yt})}-\coth{(u_{xt})}}{{\rm csch}{(u_{xy})}{\rm csch}{(u_{yt})}}\right),\\
    \frac{\partial f}{\partial u_{yt}}=\arccosh\left(\frac{\coth{(u_{xy})}\coth{(u_{xt})}-\coth{(u_{yt})}}{{\rm csch}{(u_{xy})}{\rm csch}{(u_{xt})}}\right).
 \end{array}
 \end{equation*}
 Therefore,  modulo a constant,  \eqref{f4ln} is equivalent to 
\eqref{fHH}.
It was demonstrated in \cite{XFP} that equation (\ref{c4PDE}) constitutes an alternative scalar form of the fully discrete Darboux system.
$\square$

\bigskip

\noindent{\bf Remark 4.} One can rewrite the difference equation (\ref{c4PDE}) in terms of the `geometric' variables $L_i, l_i$ so that it reduces to (\ref{ELhyphex})
in the dispersionless limit $\epsilon\to 0$.  First, we introduce $l_i$ via the relations
$$
\begin{array}{c}
\widehat{\gamma}=2\,{\rm arccoth}(\cosh{l_1})=\ln\frac{\cosh {l_1}+1}{\cosh {l_1}-1}, \\
\widehat{\beta}=2\,{\rm arccoth}(\cosh{l_2})=\ln\frac{\cosh {l_2}+1}{\cosh {l_2}-1},\\
\widehat{\alpha}=2\,{\rm arccoth}(\cosh{l_3})=\ln\frac{\cosh {l_3}+1}{\cosh {l_3}-1}.
\end{array}
$$
Using compatibility conditions of relations  (\ref{c4a9}}), we also introduce a quantity $\theta_0$ via the relations
$\triangle_x\widehat \gamma=\triangle_y\widehat \beta=\triangle_t\widehat \alpha=\frac{1}{\epsilon}\ln(1-\theta_0)$.
We take a solution of quadratic equation (\ref{q4}) in the form $ q=\frac{\chi-\sqrt{\chi^2-4\zeta}}{2}$; note that the  coefficients $\chi, \zeta$ can be expressed as in terms of $l_i$  as
$$
\chi = \frac{4\left(1-\sum_{i=1}^3 \cosh l_i \right) + \theta_0 \prod_{i=1}^3 (\cosh l_i + 1)}{\prod_{i=1}^3 (\cosh l_i - 1)},\quad  \zeta=\frac{-8}{{\prod_{i=1}^3 (\cosh l_i - 1)}}.
$$
Finally, we introduce  the notation
 $$
\theta_i = \cosh{L_i} - \frac{\theta_0}{4}  (\cosh l_i + 1) \prod_{j \neq i} \coth\frac{l_j}{2}, \quad i \in \{1, 2, 3\},
$$
where $L_i$ are related to $l_i$ via the hyperbolic cosine laws (\ref{slhyphex}), (\ref{s2hyphex}). In this notation, equation (\ref{c4PDE}), along with relations $\triangle_x\widehat \gamma=\triangle_y\widehat \beta=\triangle_t\widehat \alpha$, take the form
  \begin{equation}\label{ELhyphexdisp}
 \begin{array}{c}
\triangle_x\triangle_y{\ln\left(\theta_3+\sqrt{\theta_3^2-1+\theta_0\cosh^2\frac{l_3}{2}}\right)}+{\triangle_x\triangle_t\ln\left(\theta_2+\sqrt{\theta_2^2-1+\theta_0\cosh^2\frac{l_2}{2}}\right)}
\ \\+{\triangle_y\triangle_t\ln\left(\theta_1+\sqrt{\theta_1^2-1+\theta_0\cosh^2\frac{l_1}{2}}\right)}
-\epsilon \,\triangle_x\triangle_y\triangle_t\ln \frac{\chi+\sqrt{\chi^2-4\zeta}}{2}=0,\\
 \ \\
 \triangle_x \arccoth (\cosh  l_1)=  \triangle_y \arccoth (\cosh  l_2)=  \triangle_t \arccoth (\cosh  l_3).
 \end{array}
 \end{equation}
In the dispersionless limit $\epsilon \to 0$, we have $\theta_0\to 0$, $\theta_i\to \cosh L_i$, $\triangle_x, \triangle_y, \triangle_t \to \partial_x, \partial_y, \partial_t$, and system
(\ref{ELhyphexdisp}) goes to (\ref{ELhyphex}).

\section{Appendix}

Here we derive formulas (\ref{fS}), (\ref{fH})  and (\ref{fHH}) for the generic Lagrangian density $f$ from Section \ref{sec:Lob}.

\bigskip

\noindent {\bf Formula (\ref{fS})}:  Due to (\ref{pi}) and (\ref{sph1}), we have:
 \begin{equation*}
 \begin{array}{c}
 v_1={\rm arctanh}(\cos{a})=\frac{1}{2}\ln\frac{1+\cos{a}}{1-\cos{a}}=\frac{1}{2}\ln\frac{\cos{A}+\cos{(B-C)}}{-\cos{A}-\cos{(B+C)}}= \frac{1}{2}\ln \frac{\cos \frac{A+B-C}{2} \cos \frac{A+C-B}{2}}{\cos \frac{2\pi-A-B-C}{2}\cos \frac{B+C-A}{2}}, \\
 \\
 v_2={\rm arctanh}(\cos{b})=\frac{1}{2}\ln\frac{1+\cos{b}}{1-\cos{b}}=\frac{1}{2}\ln\frac{\cos{B}+\cos{(A-C)}}{-\cos{{B}}-\cosh{(A+C)}}= \frac{1}{2}\ln \frac{\cos \frac{A+B-C}{2} \cos \frac{B+C-A}{2}}{\cos \frac{2\pi-A-B-C}{2}\cos \frac{A+C-B}{2}}, \\
 \\
 v_3={\rm arctanh}(\cos{c})=\frac{1}{2}\ln\frac{1+\cos{c}}{1-\cos{c}}=\frac{1}{2}\ln\frac{\cos{C}+\cos{(A-B)}}{-\cos{C}-\cos{(A+B)}}= \frac{1}{2}\ln \frac{\cos \frac{B+C-A}{2} \cos \frac{A+C-B}{2}}{\cos \frac{2\pi-A-B-C}{2}\cos \frac{A+B-C}{2}}.
 \end{array}
 \end{equation*}
Rewriting $df$ in the form
$$
df=d(Av_1+Bv_2+Cv_3)-v_1dA-v_2dB-v_3dC
$$
and substituting the above expressions for $v_1, v_2, v_3$ leads, on integration, to formula (\ref{fS}).

\bigskip

\noindent {\bf Formula (\ref{fH})}:  Due to (\ref{pi1}) and (\ref{hyp1}), we have:
 \begin{equation*}
 \begin{array}{c}
 v_1={\rm arccoth}(\cosh{a})=\frac{1}{2}\ln\frac{\cosh{a}+1}{\cosh{a}-1}=\frac{1}{2}\ln\frac{\cos{A}+\cos{(B-C)}}{\cos{A}+\cos{(B+C)}}= \frac{1}{2}\ln \frac{\cos \frac{A+B-C}{2} \cos \frac{A+C-B}{2}}{\cos \frac{A+B+C}{2}\cos \frac{B+C-A}{2}}, \\
 \\
 v_2={\rm arccoth}(\cosh{b})=\frac{1}{2}\ln\frac{\cosh{b}+1}{\cosh{b}-1}=\frac{1}{2}\ln\frac{\cos{B}+\cos{(A-C)}}{\cos{{B}}+\cosh{(A+C)}}= \frac{1}{2}\ln \frac{\cos \frac{A+B-C}{2} \cos \frac{B+C-A}{2}}{\cos \frac{A+B+C}{2}\cos \frac{A+C-B}{2}}, \\
 \\
 v_3={\rm arccoth}(\cosh{c})=\frac{1}{2}\ln\frac{\cosh{c}+1}{\cosh{c}-1}=\frac{1}{2}\ln\frac{\cos{C}+\cos{(A-B)}}{\cos{C}+\cos{(A+B)}}= \frac{1}{2}\ln \frac{\cos \frac{B+C-A}{2} \cos \frac{A+C-B}{2}}{\cos \frac{A+B+C}{2}\cos \frac{A+B-C}{2}}.
 \end{array}
 \end{equation*}
Rewriting $df$ in the form
$$
df=d(Av_1+Bv_2+Cv_3)-v_1dA-v_2dB-v_3dC
$$
and substituting the above expressions for $v_1, v_2, v_3$ leads, on integration, to formula (\ref{fH}).

\bigskip

\noindent {\bf Formula (\ref{fHH})}:  Due to (\ref{hex}) and (\ref{s2hyphex}), we have:
 \begin{equation*}
 \begin{array}{c}
 v_1={\rm arccoth}(-\cosh{l_1})=\frac{1}{2}\ln\frac{\cosh{l_1}-1}{\cosh{l_1}+1}=\frac{1}{2}\ln\frac{\cosh{{L_1}}+\cosh{({L_2}-{L_3})}}{\cosh{{L_1}}+\cosh{({L_2}+{L_3})}}= \frac{1}{2}\ln \frac{\cosh \frac{{L_1}+{L_2}-{L_3}}{2} \cosh \frac{{L_1}+{L_3}-{L_2}}{2}}{\cosh \frac{{L_1}+{L_2}+{L_3}}{2}\cosh \frac{{L_2}+{L_3}-{L_1}}{2}}, \\
 \\
 v_2={\rm arccoth}(-\cosh{l_2})=\frac{1}{2}\ln\frac{\cosh{l_2}-1}{\cosh{l_2}+1}=\frac{1}{2}\ln\frac{\cosh{{L_2}}+\cosh{({L_1}-{L_3})}}{\cosh{{L_2}}+\cosh{({L_1}+{L_3})}}= \frac{1}{2}\ln \frac{\cosh \frac{{L_1}+{L_2}-{L_3}}{2} \cosh \frac{{L_2}+{L_3}-{L_1}}{2}}{\cosh \frac{{L_1}+{L_2}+{L_3}}{2}\cosh \frac{{L_1}+{L_3}-{L_2}}{2}}, \\
 \\
 v_3={\rm arccoth}(-\cosh{l_3})=\frac{1}{2}\ln\frac{\cosh{l_3}-1}{\cosh{l_3}+1}=\frac{1}{2}\ln\frac{\cosh{{L_3}}+\cosh{({L_1}-{L_2})}}{\cosh{{L_3}}+\cosh{({L_1}+{L_2})}}= \frac{1}{2}\ln \frac{\cosh \frac{{L_2}+{L_3}-{L_1}}{2} \cosh \frac{{L_1}+{L_3}-{L_2}}{2}}{\cosh \frac{{L_1}+{L_2}+{L_3}}{2}\cosh \frac{{L_1}+{L_2}-{L_3}}{2}}.
 \end{array}
 \end{equation*}
Rewriting $df$ in the form
$$
df=d(L_1v_1+L_2v_2+L_3v_3)-v_1dL_1-v_2dL_2-v_3dL_3
$$
and substituting the above expressions for $v_1, v_2, v_3$ leads, on integration, to formula (\ref{fHH}).

\section{Concluding remarks}

Here we list some further problems that may be of interest.

\begin{itemize}

\item Second-order Lagrangians from Cases 1-4  belong to the same dispersionless KP hierarchy. Similarly, their dispersive deformations belong to the full KP hierarchy.   It would be important  to merge them  into a single Lagrangian multiform in the spirit of \cite{Nijhoff2023, Nijhoff2024}.

\item It would be an interesting and nontrivial problem to classify two-field first-order integrable Lagrangians of the form
$$
\int F(v_x, v_y, v_t, w_x, w_y, w_t)\ dxdydt.
$$
A simple example thereof (related to the present paper) is the Lagrangian
$$
\int (v_x+w_x)v_yw_t\ dxdydt,
$$
with the Euler-Lagrange equations
$$
[(v_x+w_x)w_t]_y+[v_yw_t]_x=0, \qquad [(v_x+w_x)v_y]_t+[v_yw_t]_x=0.
$$
Introducing a potential $u$ such that
$$
v_yw_t=u_{yt}, \quad (v_x+w_x)w_t=-u_{xt}, \quad (v_x+w_x)v_y=-u_{xy},
$$
and taking the product of these relations, one obtains $(v_x+w_x)v_yw_t=\sqrt{u_{xy}u_{xt}u_{yt}}$, thus demonstrating the equivalence of the above two-field Lagrangian to the second-order Lagrangian from Case 1.
\end{itemize}

\bigskip

{\noindent \bf Data availability statement.}
Data sharing is not applicable to this article as no datasets were generated or analysed during the current study.

\medskip
{\noindent \bf Conflict of interest statement.}
The corresponding author states that there is no conflict of interest.

\section*{Acknowledgments}

We thank Matteo Casati for useful discussions.   The research of MVP was partially supported by the  NSFC (Grant No.12431008).

\end{document}